\documentclass[twocolumn,aps,floats,longbibliography,superscriptaddress]{revtex4-1}
\usepackage[latin9]{inputenc}
\usepackage{amsmath}
\usepackage{amssymb}
\usepackage{graphicx}
\usepackage[unicode=true,pdfusetitle,
 bookmarks=true,bookmarksnumbered=false,bookmarksopen=false,
 breaklinks=false,pdfborder={0 0 1},backref=false,colorlinks=false]
 {hyperref}
\hypersetup{
 colorlinks,linkcolor=blue,citecolor=blue,urlcolor=blue}

\makeatletter

\providecommand{\tabularnewline}{\\}

\usepackage{wasysym}

\usepackage[]{ulem}

\usepackage{xcolor}
\usepackage{pdfcolmk}
\usepackage{wasysym}

\usepackage{txfonts}

\makeatother

\begin{document}
\title{Intrinsic quantum Ising model 
 on a triangular lattice magnet TmMgGaO$_{4}$ and beyond}
\author{Changle Liu}
\thanks{These two authors contributed equally to this work.}
\affiliation{State Key Laboratory of Surface Physics and Department of Physics,
Fudan University, Shanghai, 200433, China}
\affiliation{Department of Physics and HKU-UCAS Joint Institute for Theoretical
and Computational Physics at Hong Kong, The University of Hong Kong,
Hong Kong, China}
\author{Chun-Jiong Huang}
\thanks{These two authors contributed equally to this work.}
\affiliation{Department of Physics and HKU-UCAS Joint Institute for Theoretical
and Computational Physics at Hong Kong, The University of Hong Kong,
Hong Kong, China}
\affiliation{Shanghai Branch, National Laboratory for Physical Sciences at Microscale
and Department of Modern Physics, University of Science and Technology
of China, Shanghai, 201315, China}
\affiliation{CAS Center for Excellence and Synergetic Innovation Center in Quantum
Information and Quantum Physics, University of Science and Technology
of China, Hefei, Anhui 230026, China}
\affiliation{CAS-Alibaba Quantum Computing Laboratory, Shanghai, 201315, China}
\author{Gang Chen}
\affiliation{Department of Physics and HKU-UCAS Joint Institute for Theoretical
and Computational Physics at Hong Kong, The University of Hong Kong,
Hong Kong, China}
\affiliation{State Key Laboratory of Surface Physics and Department of Physics,
Fudan University, Shanghai, 200433, China}
\affiliation{Collaborative Innovation Center of Advanced Microstructures, Nanjing
University, Nanjing, 210093, China}
\date{\today}
\begin{abstract}
The rare-earth magnet TmMgGaO$_{4}$ is proposed to be an intrinsic quantum
Ising magnet described by the antiferromagnetic transverse field Ising model (TFIM) on a triangular
lattice, where the relevant degrees of freedom are the non-degenerate
dipole-multipole doublets of the Tm$^{3+}$ ions and the transverse field
has an intrinsic origin from the weak splitting of the doublet. 
We compare this special doublet of Tm$^{3+}$ with the
dipole-octupole Kramers doublet. We study the proposed effective model
for the Tm-based triangular lattice and consider the effects 
of external magnetic fields and finite temperatures. From the ``orthogonal operator approach'',
we show that the TFIM with the three-sublattice intertwined ordered
state agrees with the experiments and further clarify the discrepancy 
in the nubmers of the magnetic sublattices and the measured magnon branches. 
We make specific predictions for the evolution of the 
magnetic properties with the external magnetic field. Furthermore,
we demonstrate that an emergent U(1) symmetry emerges in thermal
melting of the underlying orders and at the criticality, and summarize
the previously known signatures related to the finite-temperature 
Berezinskii-Kosterlitz-Thouless (BKT)
physics. We discuss the broad relevance of intrinsic quantum Ising
magnets to many other systems, especially the Tm-based materials. 
\end{abstract}
\maketitle

\section{Introduction}

Frustrated magnetism is an exciting field in modern condensed matter
physics and has been under an active investigation for the past a
few decades. Generally speaking, frustration arises from competing
interactions among local moments that cannot be satisfied simultaneously.
The strong competitions can give rise to exotic low-energy behaviors
in frustrated magnets. This feature retains in the simplest classical
antiferromagnetic Ising model, where for some particular frustrated
lattices (triangular~\cite{wannier1950antiferromagnetism}, Kagom\'{e}~\cite{kano1953antiferromagnetism},
pyrochlore~\cite{harris1997geometrical,ramirez1999zero}), there
are macroscopic degenerate ground states associated with a finite
zero-point entropy.

An interesting and important question is to consider the fate of classical
macroscopic degeneracy in presence of quantum fluctuations. Quantum
fluctuations allow tunneling within the macroscopic degenerate manifold,
therefore will lift the macroscopic degeneracy. Depending on lattice
structures, the resulting quantum ground state can be either magnetically
ordered or disordered~\cite{Moessner2000,Moessner01,hermele2004pyrochlore,pyrochlore_strain,KagomeTFIM},
owing to the so-called ``order-by-disorder'' or ``disorder-by-disorder''
mechanism~\cite{Villain1980,Shender1982,Moessner2000,PhysRevB.87.054404,PhysRevB.88.035118,PhysRevB.88.184402,PhysRevB.94.201111,PhysRevB.100.144411,PhysRevLett.122.017203}.
In practice, the simplest way to introduce quantum fluctuations is
to add a transverse field to the Ising spins. The resulting model
is the transverse field Ising model (TFIM), which has not only received
a considerable theoretical attention, but also achievable in experiments.
Moreover, this model is sign-problem free in any lattices, therefore
it can be efficiently dealt with by unbiased quantum Monte Carlo (QMC)
simulations. These qualities render TFIM a good platform for collaborations
among experimental, theoretical and numerical communities.

In realistic materials, two distinct physical origins of the transverse
field was proposed and has been summarized in Ref.~\onlinecite{Chen2019intrinsic}.
These two distinct ones are referred as extrinsic origin and intrinsic
origin. For the extrinsic origin, the transverse spin components act
as ordinary magnetic dipole moments, hence the transverse field is
directly achievable with the physical magnetic field along the transverse
directions. This mechanism applies to various Co-based Ising magnets
such as CoNb$_{2}$O$_{6}$~\cite{CoNbO,Kinross2014,Cabrera2014},
BaCo$_{2}$V$_{2}$O$_{8}$~\cite{Suga2008tomonaga,Wang2018,2018nature25466},
and SrCo$_{2}$V$_{2}$O$_{8}$~\cite{He2006,Cui2019}. For the intrinsic
origin, the transverse field is generated internally and models the
intrinsic crystal field splitting between two relevant crystal field
levels that are responsible for the low-temperature magnetism. It
was further proposed that the rare-earth magnets with low crystal
field symmetries would automatically generate such an intrinsic transverse
field for the local moments with even number of electrons. This is
because the low crystal field symmetries cannot provide enough symmetry
operations that protect the degeneracy of the crystal field levels.
Nevertheless, the intrinsic transverse field could also emerge in
the case with high crystal field symmetries. This was emphasized for
TmMgGaO$_{4}$ in the introduction of Ref.~\onlinecite{Chen2019intrinsic}
as an example of the intrinsic transverse field.

The TFIM with an intrinsic transverse field was first proposed for
TmMgGaO$_{4}$ in Ref.~\onlinecite{shen2018hidden}. 
TmMgGaO$_{4}$~\cite{Cevallos2018,YueshengTMGO,2019arXiv191202344L}
is a Mott insulator in which the Tm$^{3+}$ ions form a perfect triangular
lattice. 
Experimentally, thermodynamic~\cite{Cevallos2018,2019arXiv191202344L,YueshengTMGO} 
and detailed neutron scattering~\cite{shen2018hidden} measurements
have been performed, and the elementary spin-wave-like excitation spectrum
with respect to the magnetically ordered ground state has been 
well-recorded~\cite{shen2018hidden}.
In this system the two lowest crystal field levels of the
Tm$^{3+}$ ion that contribute to the local moment are the point-group-symmetry
demanded singlets. This intrinsic transverse field arises from the
intrinsic splitting between the two singlets. The crystal field splitting
is demanded by symmetry and appears at the atomic level, so it cannot
be ignored compared to exchange interactions and must be considered
at the first place. This is explained in details in Sec.~\ref{sec2}
and Sec.~\ref{sec3}. Moreover, in TmMgGaO$_{4}$ the transverse
and longitudinal spin components behave fundamentally different in
nature, the system exhibits antiferromagnetic dipolar order coexisting
with the preformed multipolar order due to the intrinsic transverse
field~\cite{shen2018hidden}. 
The resulting state is an example the intertwined multipolar
order, originally proposed in the context of non-Kramers doublet systems
in rare-earth magnets~\cite{GCnonK}, and also applies for TmMgGaO$_{4}$.

In this article, we systematically explore our proposed TFIM for TmMgGaO$_{4}$
and understand the physics of the Tm-based triangular lattice antiferromagnets
from a combination of techniques and perspectives that involve the
microscopics, the thermodynamic and the neutron scattering experiments,
the many-body modeling, the QMC simulation and mean-field analysis,
and the connection between the theory and the measurements. Our effort
in this work requires a sophisticated blending and a mutual feedback
amongst the microscopic physics, the many-body physics and the experimental
understanding. Therefore, this paper does not have a single thread
of logic flow in the organization of the sections. To guide the readers
well, we outline the content of the remaining parts of the papers
here. In Sec.~\ref{sec2}, we explain the non-degenerate nature of
the two lowest crystal field levels of the Tm$^{3+}$ ion and refer
them as the non-degenerate dipole-multipole doublet. We further compare
the Tm-based non-degenerate dipole-multipole doublet with the well-known
dipole-octupole doublet in Sec.~\ref{sec2b}. In Sec.~\ref{sec3},
we explore the symmetry properties of the effective spin operators
and write down the TFIM for TmMgGaO$_{4}$. In Sec.~\ref{sec4},
we provide a careful reasoning about the nature of the ground state
for TmMgGaO$_{4}$ by reading the existing experiments. This result
is independent from the microscopic modeling. If the reader is not
interested in the reasoning based on the experimental phenomena, one
can skip this section. In Sec.~\ref{sec5}, we combine mean-field
calculation, QMC simulation and theoretical arguments to establish
the finite temperature phase diagram of our proposed TFIM on the triangular
lattice. We explore the thermal BKT phase 
and transitions, as well as the emergent continuous U(1)
symmetry near the transitions. In Sec.~\ref{sec6}, we apply the
``orthogonal operator approach'' to explain the selective measurements.
From this understanding, we were able to establish the connection
between the theoretical results and the experiments. We establish
the magnetic excitations in different phases and point out the qualitative
differences between them. We clarify the the discrepancy between the
magnetic sublattices and the branches of the measured magnon excitations
in the ordered side. In Sec.~\ref{sec7}, we explore the effect of
the external magnetic fields in various physical quantities. We show
the non-monotonic behaviors of the magnetic Bragg peak in magnetic
fields, the evolution of the magnetic excitation with the fields,
and the thermodynamic behaviors. In Sec.~\ref{sec8}, we summarize
our understanding about TmMgGaO$_{4}$, and point out the relevance
of the intrinsic TFIM for other Tm-based magnets. In Appendix.~\ref{append},
we provide the results from the linear spin-wave theory where the
full structures of the magnetic excitations are available. These 
features are compared with the results from the selective measurements.

\begin{figure}[b]
\includegraphics[width=0.8\columnwidth]{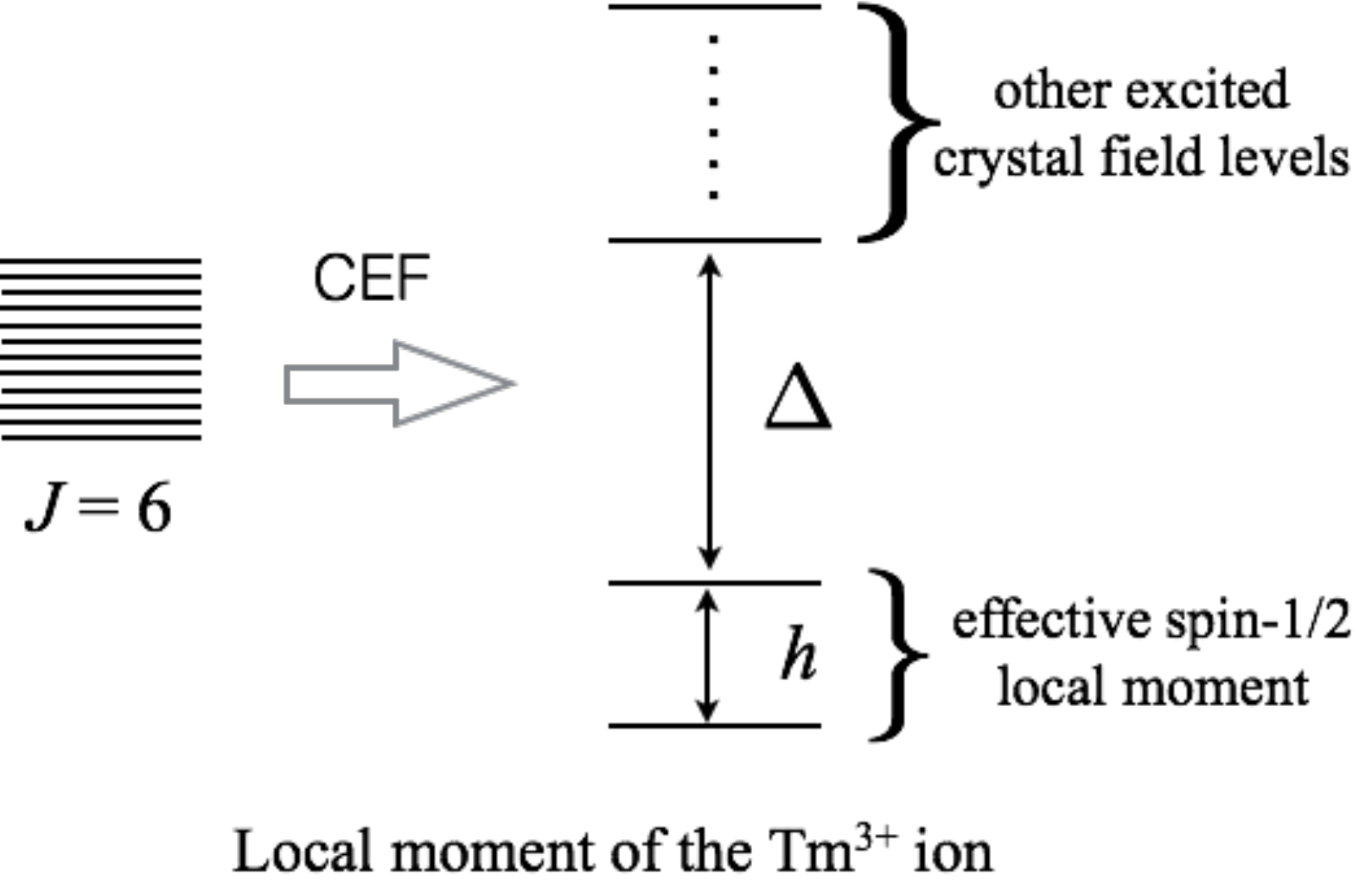} 
\caption{The splitting of the {$J=6$} total moment of the Tm$^{3+}$ ion
in TmMgGaO$_{4}$ under the D$_{3d}$ crystal electric field (CEF).
The energy separation, $h$, between the ground state singlet and
the first excited singlet is much smaller than the energy gap, $\Delta$,
to the other excited crystal field levels, and the two lowest levels
are responsible to the low-temperature magnetic properties.}
\label{cef}
\end{figure}

\section{Microscopics of $\text{TmMgGaO}{}_{4}$}

\label{sec2}

\subsection{Non-degenerate dipole-multipole doublet of Tm$^{3+}$ ion}

\begin{table}[t]
\begin{tabular}{lcc}
\hline \hline
Properties  & non-degenerate DM doublet  & DO doublet\tabularnewline
\hline 
original moment & integer  & half-odd integer\tabularnewline
time reversal  & $S^{z}\rightarrow-S^{z}$  & $S^{z}\rightarrow-S^{z}$ \tabularnewline
time reversal  & $S^{x,y}\rightarrow S^{x,y}$  & $S^{x,y}\rightarrow-S^{x,y}$\tabularnewline
degeneracy  & two separate singlets  & 2-fold degenerate \tabularnewline
3-fold rotation  & eigenvalue $+1$  & eigenvalue $-1$ \tabularnewline
\hline \hline
\end{tabular}
\caption{The comparison between the non-degenerate dipole-multipole (DM) doublet
of the Tm$^{3+}$ ion and the dipole-octupole (DO) doublet for Kramers
ions. }
\end{table}

Here, for the purpose of completeness, we explain this microscopic
physics of the Tm$^{3+}$ ion in the language that is aligned with
our early works in the field, and compare it with the well-known dipole-octupole
doublet for half-integer moments. The non-degenerate dipole-multipole
doublet nature of the Tm$^{3+}$ ion in $\text{TmMgGaO}{}_{4}$ was
clarified and carefully modelled in Ref.~\onlinecite{shen2018hidden}.
The Tm$^{3+}$ ion has a total orbital angular moment ${L=5}$ and
total spin moment ${S=1}$, and the spin-orbit coupling gives a total
moment ${J=6}$~\cite{shen2018hidden,YueshengTMGO}. The thirteen-fold
degeneracy of the total moment is further split by the crystal field.
Unlike the usual degeneracy for Kramers doublets and non-Kramers doublet,
the ground state and the first excited state of the Tm$^{3+}$ ion
are both singlets (see Fig.~\ref{cef}). They are \textsl{not degenerate},
and there is no reason to support the degeneracy of these two states.
Each state in the relevant quasi-doublet is an one-dimensional irreducible
representation of the D$_{3d}$ point group, and there should always
be a crystal field splitting between two states of the quasi-doublet.
This crystal field splitting was further modeled as an intrinsic transverse
field by us in Ref.~\onlinecite{shen2018hidden}. All these are observed
from the form of the wavefunction for each state in the quasi-doublet.
The wavefunction is a linear superposition of $|{J^{z}=3n}\rangle$
where $n$ is an integer and $J^{z}$ is defined on the local 3-fold
rotational axis of the triangular lattice. In terms of the notation
in Ref.~\onlinecite{shen2018hidden}, the two wavefunctions are 
\begin{eqnarray}
|\Psi_{g}\rangle & = & c_{6}[|{6}\rangle+|{-6}\rangle]+c_{3}[|{3}\rangle-|{-3}\rangle]+c_{0}\,|{0}\rangle,\\
|\Psi_{e}\rangle & = & c_{6}'[|{6}\rangle-|{-6}\rangle]+c_{3}'[|{3}\rangle+|{-3}\rangle],
\end{eqnarray}
where $|3n\rangle$ (with $n\in\mathbb{Z}$) refers to the quantum
number of $J^{z}$, $|\Psi_{g}\rangle$ ($|\Psi_{e}\rangle$) refers
to the ground state (the first excited crystal field level), and the
two singlets carry $A_{1g}$ and $A_{2g}$ representation of the $D_{3d}$
point group, respectively. Here $c_{6}^ {},c_{3}^ {},c_{0}^ {},c_{6}',c_{3}'$
are real numbers with $|c_{6}^ {}|\approx|c_{6}'|\gg c_{3}^ {},c_{3}',c_{0}^ {}$.
Their nature of the one-dimensional irreducible representation can
be simply seen by applying the three-fold rotation operation, 
\begin{eqnarray}
e^{-i\frac{2\pi}{3}J^{z}}|{J^{z}={3n}}\rangle=|{J^{z}={3n}}\rangle,
\end{eqnarray}
other integer spin numbers do not have this property, and they often
give rise to two-dimensional representation of the D$_{3d}$ point
group. The point group symmetry does not allow the degeneracy between
the ground state singlet and the first excited singlet. Due to the
intrinsic integer spin (${J=6}$) in nature for the Tm$^{3+}$ ion,
there is no Kramers' theorem's protection, either.

It is ready to notice that both $|\Psi_{g}\rangle$ and $|\Psi_{e}\rangle$
are non-magnetic, and thus thinking locally about the single-ion physics
would not lead to any magnetism. The magnetism should come from the
exchange interaction between the local moments. The intrinsic competition
between the single-ion physics and the exchange interaction is captured
and modelled as an intrinsic TFIM by us~\cite{shen2018hidden} and
will be explained in great details in Sec.~\ref{sec3}.

As the Tm$^{3+}$ doublet in this context was sometimes referred as
a non-Kramers doublet, we here clarify their difference. The usual
non-Kramers doublet, that occurs in for example the Pr$^{3+}$ ion~\cite{PhysRevB.94.134428,PhysRevB.94.205107}
of Pr$_{2}$Zr$_{2}$O$_{7}$ and Pr$_{2}$Ir$_{2}$O$_{7}$ or other
rare-earth triangular lattice magnets~\cite{GCnonK}, is composed
of two degenerate crystal field states, and their degeneracy is not
protected by time reversal but protected by the point group symmetry.
These states comprise the \textsl{two-dimensional} irreducible representation
of the point group symmetry. In comparison, the Tm$^{3+}$ doublet
is two non-degenerate point-group singlets that are two independent
\textsl{one-dimensional} irreducible representations.

\subsection{Comparison with dipole-octupole doublet}

\label{sec2b}

It is instructive to compare the non-degenerate dipole-multipole doublet
of the Tm$^{3+}$ ion with the dipole-octupole doublet that also arises
from the one-dimensional irreducible representations of the D$_{3d}$
point group. The dipole-octupole doublet was first introduced in the
context of pyrochlore magnets in Refs.~\onlinecite{GC_DO,Chen2017symm}
and then extended to the triangular lattice magnets in Refs.~\onlinecite{GCoctu,Yaodong1,GCnonK}.
The dipole-octupole doublet was found to be applicable to the Nd$^{3+}$
ion in various Nd-based pyrochlores~\cite{Lhotelnd2015,Anand2015,Bertin2015,Xu2015,Hatnean2015,Petit2016observation,Benton2016,Dalmas2017},
the Sm$^{3+}$ ion in Sm$_{2}$Ti$_{2}$O$_{7}$~\cite{Mauws2018,Pecanha2019},
the Ce$^{3+}$ ion in Ce$_{2}$Sn$_{2}$O$_{7}$~\cite{Sibillece2015,Sibille2019quantum,Chen2017symm}
and Ce$_{2}$Zr$_{2}$O$_{7}$~\cite{Gaudet2019,Gao2019,2019arXiv190207075L},
and the Er$^{3+}$ ion in the spinel compounds~\cite{Lago2010,Gao2018dipolar}.
In fact, the dipole-octupole doublet can broadly exist in magnets
whose local environment has a D$_{3d}$ point group symmetry. In this
regards, other lattice geometry such as honeycomb magnet could support
the dipole-octupole doublet~\cite{unpub}. As a parallel thought,
the Tm$^{3+}$ dipole-multipole doublet could broadly exist in many
other structures. This is discussed in some details in Sec.~\ref{sec8}.

For the dipole-octupole doublet, the wavefunction of each state in
the doublet is a linear superposition of $|{J^{z}=3n/2}\rangle$ where
$n$ is an odd integer and $J^{z}$ is defined on the local 3-fold
rotational axis. On the triangular lattice, the local 3-fold rotational
axis aligns with the global $z$ axis, while it is not the case for
the pyrochlore lattice. The reason that it is a one-dimensional irreducible
representation can be seen by applying the three-fold rotation operation,
\begin{eqnarray}
e^{-i\frac{2\pi}{3}J^{z}}|{J^{z}=\frac{3n}{2}}\rangle=-|{J^{z}=\frac{3n}{2}}\rangle.
\end{eqnarray}
The eigenvalue of the 3-fold rotation is $-1$, instead of $+1$ for
the dipole-multipole doublet of the Tm$^{3+}$ ion. Unlike the non-degenerate
dipole-multipole doublet of the Tm$^{3+}$ ion, the dipole-octupole
doublet for the Kramers ion is degenerate, and the degeneracy is protected
by the time reversal symmetry due to the Kramers' theorem for the
half-integer spin moment.

\section{Effective model of $\text{TmMgGaO}{}_{4}$}

\label{sec3}

\begin{figure}[t]
\label{Fig:tri}\includegraphics[width=1\columnwidth]{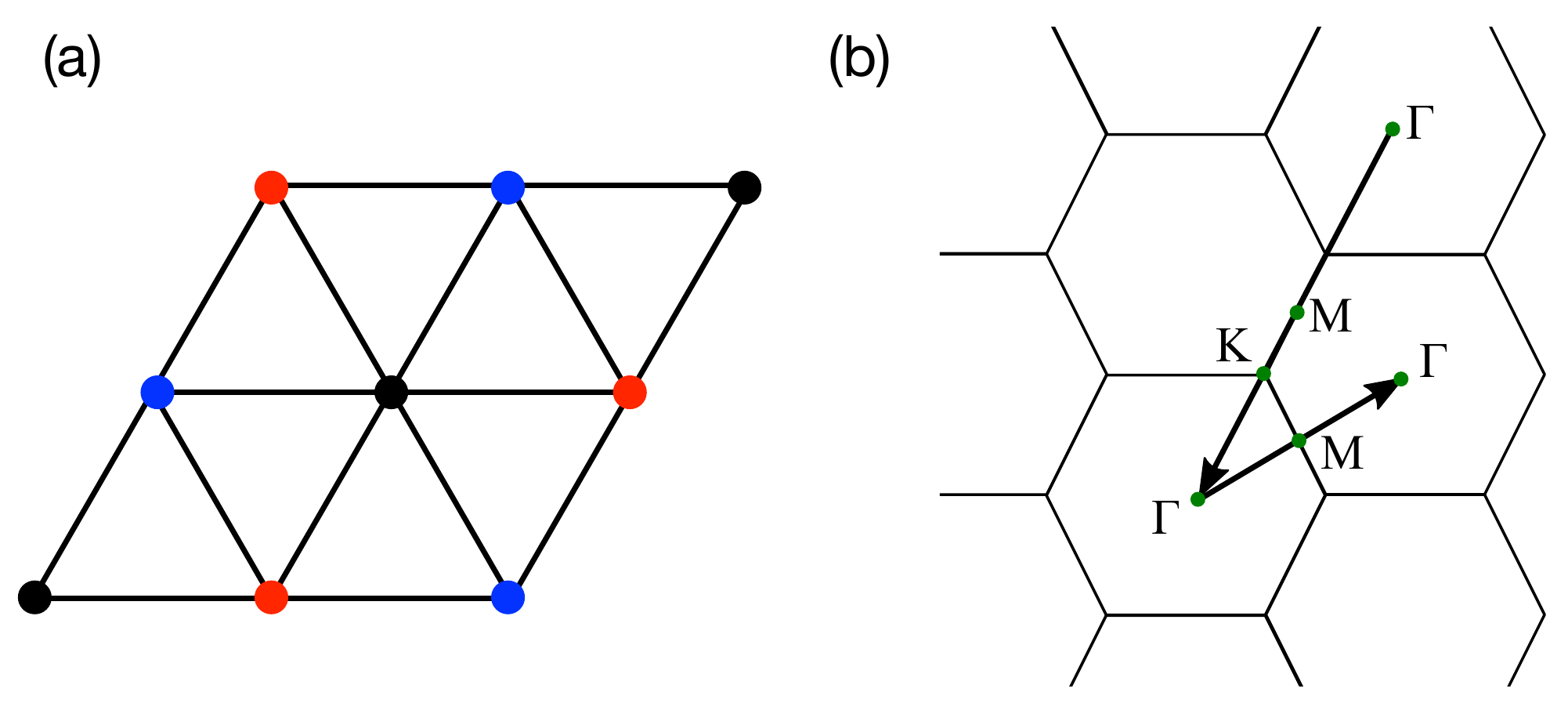} \caption{(a) Definition of the triangular lattice. The three sublattices are
marked by blue, red and black colors, respectively. (b) The Brillouin
zone of triangular lattice.}
\end{figure}

Like any two-level systems, the Tm$^{3+}$ doublet can be captured
by an effective spin-1/2 operators that operate on the manifold of
the doublet. We define the following effective spin-1/2 operator $\boldsymbol{S}_{i}$
on each Tm site, 
\begin{align}
S_{i}^{x} & =\frac{i}{2}\left(|\Psi_{i,e}\rangle\langle\Psi_{i,g}|-|\Psi_{i,g}\rangle\langle\Psi_{i,e}|\right),\\
S_{i}^{y} & =\frac{1}{2}\left(|\Psi_{i,g}\rangle\langle\Psi_{i,g}|-|\Psi_{i,e}\rangle\langle\Psi_{i,e}|\right),\\
S_{i}^{z} & =\frac{1}{2}\left(|\Psi_{i,g}\rangle\langle\Psi_{i,e}|+|\Psi_{i,e}\rangle\langle\Psi_{i,g}|\right).
\end{align}
We can see from the effective spin definition that $|\Psi_{i}^{\pm}\rangle$
are eigenstates of $S^{y}$ with eigenvalue ${S^{y}=\pm1/2}$, while
the $S^{x}$ and $S^{z}$ components introduces hybridization between
$|\Psi_{i}^{\pm}\rangle$. From our definition of the spin operators,
the point group symmetry demanded splitting between $|\Psi_{i,g}\rangle$
and $|\Psi_{i,e}\rangle$ is modelled as an intrinsic transverse field
on the $y$ component of the effective spin, i.e. $-h\sum_{i}S_{i}^{y}$,
where $h$ is the crystal electric field splitting. Moreover, the
``$x$'' and ``$y$'' in $S^{x}$ and $S^{y}$ are defined in
the internal Hilbert space of the crystal field states, $|\Psi_{i,g}\rangle$
and $|\Psi_{i,e}\rangle$, and have no connection to the real space.
However, we often refer these two components as ``in-plane components''
for convenience. The $S^{z}$ component has its physical meaning both
for the real space and for the internal Hilbert space.

It is illuminating to obtain the symmetry properties of the effective
spin operators. Under the point group symmetry and the time reversal
(${\mathcal{T}}$) operations, the effective spin components transform
as, 
\begin{align}
C_{3}: & \quad S_{i}^{x}\rightarrow+S_{i}^{x},S_{i}^{y}\rightarrow S_{i}^{y},S_{i}^{z}\rightarrow+S_{i}^{z},\\
C_{2}': & \quad S_{i}^{x}\rightarrow-S_{i}^{x},S_{i}^{y}\rightarrow S_{i}^{y},S_{i}^{z}\rightarrow-S_{i}^{z},\\
{\mathcal{T}}: & \quad S_{i}^{x}\rightarrow+S_{i}^{x},S_{i}^{y}\rightarrow S_{i}^{y},S_{i}^{z}\rightarrow-S_{i}^{z},
\end{align}
where $C_{3}$ refers to three-fold rotation about c axis, $C_{2}'$
refers to two-fold rotation along the bond direction. The in-plane
component $S^{x}$ and $S^{y}$ operators are time reversal even and
transform as even-order multipole moments under crystal symmetries.
From the wavefunction of $|\Psi_{g}\rangle$ and $|\Psi_{e}\rangle$,
it is clear that $S^{x}$ and $S^{y}$ mostly connect $|{J^{z}=6}\rangle$
and $|{J^{z}=-6}\rangle$ and are mostly involve the 12th order multipole
moments. The $S^{z}$ component is odd under time reversal and transforms
as dipole moment. The low-temperature magnetization is provided by
$\langle\sum_{i}S_{i}^{z}\rangle$. While the dipole moment, $S^{z}$,
can be probed by neutron scattering, the multipole moments are hidden
or invisible in most conventional experimental probes and are often
referred as ``hidden orders'' or ``hidden components'' in the
literature.

Based on the saturated values of the magnetic moment in the field,
one can infer that the Tm local moment is almost an Ising spin. This
is also understood from the wavefunctions of $|\Psi_{g}\rangle$ and
$|\Psi_{e}\rangle$ where $|{J^{z}=\pm6}\rangle$ are dominant. The
exchange interaction between the Tm local moments would be primarily
an Ising interactions. The exchange interaction between the transverse
components are strongly suppressed as $S^{x}$ and $S^{y}$ are high
order multipole moments and they are even higher than the quadrupole
moments. The resulting effective Hamiltonian for the interacting Tm
local moment is the TFIM 
\begin{equation}
H=\sum_{\langle ij\rangle}J_{zz}S_{i}^{z}S_{j}^{z}-\sum_{i}\left(hS_{i}^{y}+BS_{i}^{z}\right),
\label{eq:ham}
\end{equation}
where ${B\equiv\mu_{B}g_{\parallel}B^{z}}$ represents the external
magnetic field along the $z$ direction, and $h$ is the intrinsic
transverse field. In Ref.~\onlinecite{shen2018hidden}, we actually
included a tiny second neighbor Ising interaction $J_{2}$ to improve
the fitting to the experiments. As the interaction energy scale between
the Tm local moment is already quite small, the tiny $J_{2}$ does
not change the qualitative physics in this paper. Thus, we will rely
on the above minimal model to capture the essential physics about
TmMgGaO$_{4}$ and other intrinsic quantum Ising magnets.
Nevertheless, if one is more interested in the quantitative aspects,
other non-essential and non-universal ingredients should be included into our Hamiltonian. 
These would involve the long-range dipole-dipole ($S^z$-$S^z$) interaction 
and the van-Vleck process through the excited crystal field states. 

The magnetic moment of the Tm$^{3+}$ ion is much larger than the one for 
the Yb$^{3+}$ ion in YbMgGaO$_4$. 
Although the first-neighbor dipole-dipole interaction may be 
incorporated with the superexchange interaction and modelled as
a total $J_{zz}$ interaction, we here estimate the further neighbor
dipole-dipole interaction and find that the 
second-neighbor dipole-dipole interaction is 0.48K, the 
third-neighbor dipole-dipole interaction is 0.31K,
the fourth-neighbor dipole-dipole interaction is 0.134K,
the fifth-neighbor dipole-dipole interaction is 0.092K,
and the sixth-neighbor dipole-dipole interaction is 0.053K. 
If one simply attributes all the further neighbor interactions 
beyond the first neighbor to the dipole-dipole interactions, 
one readily finds from the Curie-Weiss temperature ($-19$K)~\cite{shen2018hidden} 
that the first neighbor interaction $J_{zz}$ is ${\sim11.5}$K and should 
dominate over further neighbor interactions. Thus, it is legitimate for us 
to keep only the first neighbor or first few neighbor interations in TRIM
to capture the qualitative physics. 
Moreover, the energy gap from the doublet to the lowest crystal 
field excited level is smaller than 
the one in YbMgGaO$_4$~\footnote{Yao Shen, Private communication.}. 
Thus the virtual van-Vleck process 
could further bring extra ingredients into the quantitative modelling.

\section{Qualitative understanding from experiments}
\label{sec4}

While the intrinsic quantum Ising model for TmMgGaO$_{4}$ was derived
from the microscopics in Sec.~\ref{sec3} and in Ref.~\onlinecite{shen2018hidden},
various physical insights can be gained from the careful reading of
the existing experiments before the derivation and solving of this
model. To the best of our knowledge, the single crystal sample of
TmMgGaO$_{4}$ and its basic structure and thermodynamic properties
were reported in Ref.~\onlinecite{Cevallos2018}. Even though the
measurements were performed above 1.8K, the magnetization results
already show the strong Ising-like features. More low-temperature
thermodynamic measurements were obtained in Ref.~\onlinecite{YueshengTMGO},
and the results were interpreted from classical Ising moments with
competing Ising interactions. The low-temperature magnetic state was
suggested to be a stripe order with an alternating Ising spin arrangement
on two magnetic sublattices, and the transition to the stripe order
was suggested to occur at $\sim0.27$~K. This spin state has an ordering
wavevector at the momentum point $M$ in the Brillouin zone. The detailed
elastic and inelastic neutron scattering measurements were performed
in Ref.~\onlinecite{shen2018hidden} together with the low-temperature
thermodynamic measurements. The appearance of the magnetic Bragg peak
at the wavevector $K$ coincides with the peak at $\sim1$~K in the
specific heat data. The ordering wavevector $K$ indicates a three-sublattice
magnetic order structure, which differs from the proposal of stripe
order in Ref.~\onlinecite{YueshengTMGO}. Moreover, the data-rich inelastic
neutron scattering measurements show a coherent spin-wave like excitation
spectrum with a well-defined dispersion.

The first question is, what does neutron scattering measurement actually
detect? This question is also useful in our actual calculation of
the physical properties. There are two ways to think about this. The
first way is to rely on experiments, i.e., using experiments to understand
experiments. The magnetization measurements suggest that the in-plane
components of the local moment, if they exist, almost do not respond
to the application of the external magnetic field. The neutron spin
couples to the local moment in the same way as the external magnetic
field would do. The neutron spin naturally picks up the out-of-plane
component, $S^{z}$, of the local moment. Thus, the magnetic Bragg
peak at the $K$ point indicates a three-sublattice structure for the
$S^{z}$ components. Likewise, the inelastic neutron scattering detects
the dynamic part of the $S^{z}$-$S^{z}$ correlator, even though
it is a regular neutron scattering measurement and functions as a
polarized neutron scattering. The second way is based on the microscopics.
Microscopic analysis tells us that, the out-of-plane component, $S^{z}$,
is a magnetic dipole moment, and couples linearly with the external
magnetic field, while the in-plane components, $S^{x,y}$, are the
magnetic multipole moments and do not couple to the magnetic field
at the linear order. The coupling of $S^{x,y}$ to the field could
occur at high orders but is suppressed due to the large crystal field
energy separation between the non-degenerate dipole-multipole doublet
and the other highly excited crystal field levels. Knowing the microscopic
facts, one can immediately conclude that the neutron scattering measurement
detects the properties of the $S^{z}$ component.

The next level of question is how to reconcile these experiments.
Again, we first rely on the experiments and then turn to the microscopics.
Let us start with the first possibility. If the Tm$^{3+}$ local moment
is truly Ising spin with Ising interaction like the one used in 
Ref.~\onlinecite{YueshengTMGO}, then there is no quantum mechanics, 
and there should not be any dispersion-like excitation. 
It is not the case in the inelastic neutron measurement
in Ref.~\onlinecite{shen2018hidden}. The second possibility is that
the local moment is a quantum spin and all the three components are
present and active in the physical Hilbert space. From the elastic
neutron scattering measurement, we can conclude that, the $S^{z}$
component has ${\langle S^{z}\rangle\neq0}$ and develops a three-sublattice
structure at low temperatures, but we do not know anything about the
in-plane components $S^{x}$ and $S^{y}$. 
If ${\langle S^{x}\rangle=\langle S^{y}\rangle=0}$,
then the dynamic correlation of $S^{z}$-$S^{z}$, that is detected
by the inelastic neutron scattering measurement, would simply be a
two-magnon continuum. This is again not the case in the experiments.
Thus, we expect that, at least one component of the in-plane components
should be non-zero, even though they are not experimentally visible.
The role of the $S^{z}$ operator is to flip the in-plane component
and create a coherent magnetic excitation. To summarize this part
of reasoning, we conclude from reading the experiments with an intertwined
dipolar and multipolar orders for the ground state of TmMgGaO$_{4}$,
\begin{eqnarray}
 &  & \langle S^{z}\rangle\neq0\text{\quad with a three-sublattice structure},\\
 &  & \langle S^{x}\rangle\neq0\text{\quad and/or \quad}\langle S^{y}\rangle\neq0.
\end{eqnarray}
From the microscopics and our modeling, it is obvious to see that
the invisible component, $S^{y}$, is non-zero as it is polarized
by the intrinsic transverse field.

The final issue to resolve is to see whether our microscopic modeling
can provide useful understanding of the physical properties and insights
for future experiments on TmMgGaO$_{4}$ and/or other Tm-based triangular
lattice magnets. This is carried out in the next few sections.

\section{Phase diagram}

\label{sec5}

\subsection{Mean-field analysis}

\begin{figure}
\includegraphics[width=1\columnwidth]{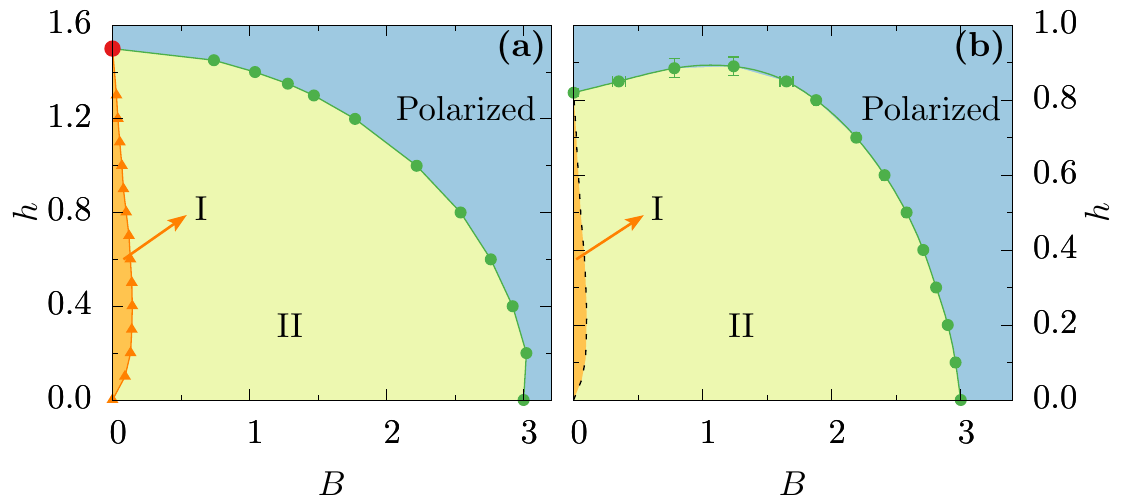} 
\caption{Phase diagram of the model in Eq.~\eqref{eq:ham}. Here we set the
energy unit ${J_{zz}=1}$. Two three-sublattice ordered phases, I
(orange) and II (green) and a polarized phase are found in the phase
diagram. The red dot represents quantum phase transition with (2+1)d
XY universality class. The left (a) is the mean-field result, and
the right (b) is the Monte Carlo result that is calculated at inverse
temperature $\beta=8L$ with system sizes $N=L\times L$ $(L=6,12,24)$.
The phase boundary from phase I to II is difficult to be captured
in the current algorithm and are schematic here.
}
\label{Fig:phases} 
\end{figure}

The TFIM on the triangular lattice has been well-studied in the absence
of the external magnetic field~\cite{Moessner01,TFIM_MonteCarlo},
while the situation with the longitudinal field has not been investigated
yet. To gain some physical insight into the ground state phase diagram,
we first tackle with the Weiss mean-field approximation by decoupling
interactions between different spins as 
\begin{equation}
S_{i}^{z}S_{j}^{z}\rightarrow\langle S_{i}^{z}\rangle S_{j}^{z}+S_{i}^{z}\langle S_{j}^{z}\rangle-\langle S_{i}^{z}\rangle\langle S_{j}^{z}\rangle.
\end{equation}
Here the mean-field order parameter $\langle S_{i}^{z}\rangle$ needs
to be solved self-consistently. The mean-field phase diagram is depicted
in Fig.~\ref{Fig:phases}(a).

In the Ising limit (without the transverse and longitudinal fields),
the system lies at a classically critical state that hosts a macroscopic
ground-state degeneracy: any spin configuration with ``2-up-1-down''
or ``1-up-2-down'' has the minimal energy. With introducing the
transverse field $h$, quantum fluctuations allow quantum tunneling
within the massively degenerate manifold. This quantum tunneling lifts
the macroscopic degeneracies and eventually stabilizes a three-sublattice
long-range ordered phase (dubbed the three-sublattice ``I'' state)
as the ground state owing to the quantum order-by-disorder mechanism.
Since the three-sublattice ordering is entirely contributed by the
quantum fluctuations, it is relatively weak and is controlled by quantum
fluctuation $h$ in a \textsl{non-monotonic} fashion: with $h$ being
too small the quantum order-by-disorder effect is weak, while for
a very large $h$ the polarization effect becomes more important,
suppresses the three-sublattice ordering and drives the system into
the ``quantum disordered'' state where the spins are fully polarized
along the transverse direction. Although the above results are obtained
mean-field level, they are consistent with those obtained via quantum
dimer model mapping where quantum fluctuations are taken into account
in a perturbative manner~\cite{Moessner2000,Moessner01}.

As the external longitudinal field $B$ is applied at the Ising limit,
the system immediately becomes unstable against the magnetic ordering
due to the criticality at this point. The resulting state is another
three-sublattice ordered state called ``1/3-plateau'' state with
a ``2-up-1-down'' structure on each triangular plaquette. Unlike
the pure quantum origin in the ``I'' phase, the three-sublattice
ordering of the plateau state arises at the classical level and are
more stable. The plateau state remains as the ground state upon increasing
the magnetic field until the system becomes fully polarized at ${B_{c}=3J_{zz}}$
through a first-order transition. When the quantum fluctuation $h$
is switched on, the three-sublattice ``plateau'' state becomes the
``quasi-plateau'' phase (dubbed three-sublattice ``II'' state)
because the total magnetization is no longer a good quantum number.
Moreover, as the three-sublattice ``I'' phase is generated by the
quantum fluctuations and is {\sl fully gapped}, 
it is stable against the weak perturbations.
But since that the ordering is rather weak, a small external field
$B$ could drive the system to the quasi-plateau state across a phase
transition. The transition from ``I'' to ``II'' state is of the
second-order, while the transition from ``II'' to the fully polarized
state is of the first-order, consistent with what happens at ${h=0}$
limit. The two phase boundaries terminate at the classical critical
point ${h=0}$, and at the quantum critical point ${h_{c}^{\textrm{MF}}=1.5J_{zz}}$,
both located along ${B=0}$ axis. These are depicted in Fig.~\ref{Fig:phases}(a)
and obtained from the mean-field analysis.

\subsection{Path-integral quantum Monte Carlo method}

To examine our mean-field results, we perform the quantum Monte Carlo
(QMC) simulations. We choose the the path-integral with the $\{S_{i}^{z}\}$
basis. The partition function of the original model is mapped onto
a worldline representation: 
\begin{eqnarray}
{\mathcal{Z}} & = & \text{Tr}\left[e^{-\beta\mathcal{H}}\right]
=\sum\limits _{\{\alpha_{0}\}}\langle\alpha_{0}|e^{-\beta\mathcal{H}}|\alpha_{0}\rangle\nonumber \\
 & = & \lim_{{\scriptstyle {{d\tau=\frac{\beta}{n}}\atop {\scriptstyle {n\rightarrow\infty}}}}}\sum\limits _{{\scriptstyle {{\{\alpha\}}\atop {\scriptstyle {\alpha_{n}=\alpha_{0}}}}}}\langle\alpha_{n}|e^{-\mathcal{H}d\tau}|\alpha_{n-1}\rangle\cdots\langle\alpha_{1}|e^{-\mathcal{H}d\tau}|\alpha_{0}\rangle\nonumber \\
 & = & \sum\limits _{\{\alpha\}}\sum_{k=0}^{\infty}\int_{0}^{\beta}\cdot\cdot\int_{\tau_{_{2k-1}}}^{\beta}\prod_{i=1}^{2k}d\tau_{i}\ h^{2k}e^{{-\int_{0}^{\beta}U(\tau)d\tau}},
 \label{eq:parfun}
\end{eqnarray}
where 
\begin{eqnarray}
U(\tau)=\langle\alpha(\tau)|\left(\sum_{\langle ij\rangle}J_{zz}S_{i}^{z}S_{j}^{z}-\sum_{i}BS_{i}^{z}\right)|\alpha(\tau)\rangle.
\end{eqnarray}

\begin{figure}[t!]
\centering \includegraphics[width=1\linewidth]{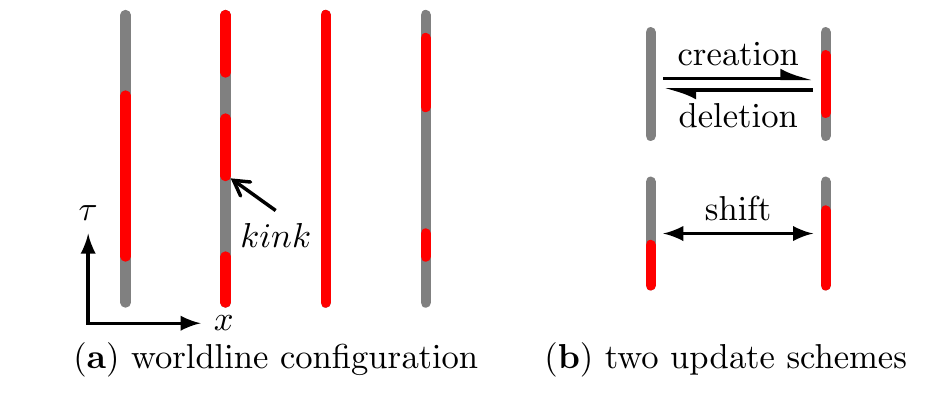} \caption{\label{fig:worldline}The worldline configuration under imaginary
time evolution and update schemes. (a) is a worldline configuration
of four spins in a chain. Different colors correspond to different
spin $S^{z}$ states. Along the imaginary time, every spin worldline
may be divided into several flats by cause of $S_{i}^{y}$. (b) the
diagrammatic sketch of the update schemes.}
\end{figure}

In Fig.~\ref{fig:worldline}(a), we depict a representative worldline
configuration that contributes to the partition function. The transverse
field term of $\sum_{i}hS_{i}^{y}$ causes the spin $S^{z}$ to flip,
and we refer such a flipping event as a kink. The temporal periodic
boundary condition ${|\alpha(0)\rangle=|\alpha(\beta)\rangle}$ of
the path integral demands the number of the kinks $N_{k}$ to be even
with ${N_{k}=2k}$ ($k\in\mathbb{Z}$) in Eq.~\eqref{eq:parfun}.
Due to the presence of the longitudinal field $B$, the cluster update
fails, and instead, we design a metropolis algorithm that contains
two update schemes, creation/deletion flat and shift kink, as shown
in Fig.~\ref{fig:worldline}. The calculations of acceptance rates
of update schemes are quite standard through the detailed balance
equation, and we will not show them explicitly here. The thermal annealing
procedure is employed to deal with the freezing issue of the Monte
Carlo simulation.

In the QMC simulations, we take the system sizes $N=L\times L$ $(L=6,12,24)$
with periodic boundary condition. The ground-state phase diagram is
calculated at inverse temperature $\beta=8L$ and the result is shown
in Fig.~\ref{Fig:phases}(b) through the finite size scaling. The
QMC phase diagram agrees with the mean-field one at the qualitative
level. The locations of the phase boundaries differ quantitatively.
The critical field 
${h_{c}^{\textrm{MC}}\approx0.82J_{zz}}$~\cite{TFIM_MonteCarlo,Wang2017}
is almost half of the mean-field result ${h_{c}^{\textrm{MF}}=1.5J_{zz}}$
with zero external field ${B=0}$. This is as expected, as the mean-field
approximation underestimates the quantum fluctuations especially for
the phase boundaries. Nevertheless, the mean-field theory provides
the essential physical understanding and insights for the magnetic
properties of the system.

\subsection{Finite temperature regimes and BKT transitions}

In this subsection, we extend the analysis from the zero temperature
or the near-ground-state low temperatures to finite temperatures and
study the finite temperature properties and the phase transitions
out of the ordered one. To reveal the finite-temperature transitions,
it is necessary to perform the field theoretical analysis near the
transition and then supplement with the QMC calculations. The three-sublattice
order parameter is characterized by the Fourier transformed $S^{z}$
dipolar component at the $K$ point. This can be captured by the following
complex field 
\begin{equation}
\psi=\frac{1}{\sqrt{3}}(m_{1}+m_{2}e^{i2\pi/3}+m_{3}e^{-i2\pi/3}),\label{eq:op}
\end{equation}
where $m_{i}$ (${i=1,2,3}$) are the dipolar magnetizations of the
three sublattices at the neighboring sites, and we have set the lattice
constant to unity. We can see that $\psi$ characterizes the three-sublattice
ordering, as ${\psi=0}$ occurs only when ${m_{1}=m_{2}=m_{3}}$, where
the three-sublattice order vanishes. The transformation of the field
variable $\psi$ under the lattice translation $T_{\hat{x}}$ and
the time reversal ${\mathcal{T}}$ operation take the following form
\begin{align}
T_{\hat{x}}: & \psi\rightarrow\psi e^{i2\pi/3},\\
{\mathcal{T}}: & \psi\rightarrow-\psi.
\end{align}

\begin{figure}
\includegraphics[width=1\columnwidth]{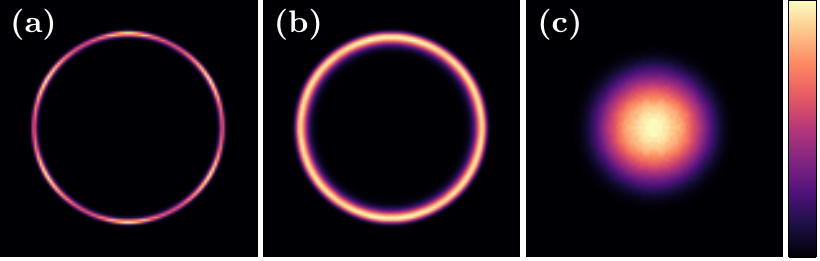} 
\caption{Histograms of the order parameter $\psi$ in different temperature
regimes obtained from QMC simulations. (a): three-sublattice long-range
ordered state at low temperature ${T=0.025}$. (b): quasi-long-range
ordered ``BKT phase'' at intermediate temperature ${T=0.075}$. (c):
 disordered state at high temperature ${T=0.225}$. In the QMC simulations
we set the model parameter ${J_{zz}=1}$, ${h=0.4}$, ${B=0}$ and the
system size ${L=12}$.}
\label{Fig:histogram} 
\end{figure}

For the three-sublattice ``I'' state, the spin alignments at the
three sublattices are different from one another, therefore the ground
state is six-fold degenerate. With zero external magnetic field, the
$\psi$ corresponding to the ground states are located at a circle
in the complex plane with $\mathrm{Arg}\,\psi=(2n+1)\pi/6$ $(n=0,1,...,5)$
that are protected by the translation and time-reversal symmetry (see
Fig.~\ref{Fig:histogram}(a)). This clock anisotropy is robust against
the short-range interactions such as weak transverse exchange and
next-nearest-neighbor Ising interactions that are present in the materials,
therefore our analysis remains valid against these perturbations.
In the vicinity of the melting of the magnetic order, the coarse-grained
Landau-Ginzburg-Wilson free energy dictates the $\mathbb{Z}_{6}$
clock anisotropy takes the following form~\cite{TFIM_OP} 
\begin{align}
H_{\textrm{LGW}}= & -K|\nabla\psi|^{2}+r\psi^{*}\psi
+u_{4}(\psi^{*}\psi)^{2}+u_{6}(\psi^{*}\psi)^{3}
\nonumber \\
 & +v_{6}(\psi{}^{6}+\psi^{*}{}^{6})
\label{eq:EFT}
\end{align}
with ${\psi=|\psi|e^{i\theta}}$, where $\theta$ corresponds to 
the phase of the field $\psi$. The $\mathbb{Z}_{6}$ clock
anisotropy term $v_{6}$ has a significant implication on the nature
of thermal and quantum phase transitions. First of all, let us examine
the thermal melting of the three-sublattice states. Since the clock
anisotropy term is brought about by the quantum fluctuations from
the transverse field and is expected to be small, the phase fluctuations
of the order parameter $\psi$ is soft therefore becomes important
for the thermal melting at the first stage. By integrating out the
amplitude fluctuations, we obtain the 2D XY model with a $\mathbb{Z}_{6}$
clock anisotropy. This theory exhibits an approximate 
self-duality~\cite{Chen_2017,2012NuPhB.854.780O},
where the dual theory is described in terms of vortices of $\theta$
that acts as the disorder parameter of the original theory. It was
previously understood that, certain self-dual quantum critical points
can put constraints on the physical observables such as a non-divergent
Gr\"{u}neisen ratio~\cite{PhysRevLett.123.230601}. The current transition
is an approximate self-duality and is driven by temperature. Whether
an analogous property can occur here will be explored in future work.

\begin{figure}
\includegraphics[width=0.9\columnwidth]{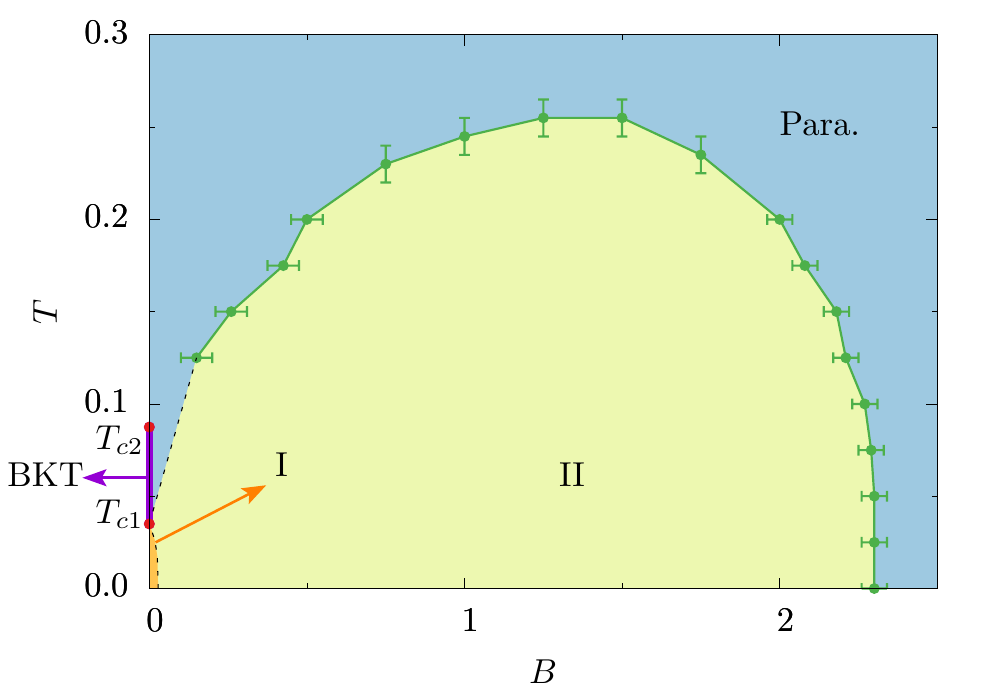} 
\caption{Finite temperature phase diagram with an external magnetic field $B$
obtained from QMC simulations. The parameter we take is $J_{zz}=1$,
$h=0.65$ where the ground state is a three-sublattice ordered state.
The ``BKT phase'' at $B=0$ axis is marked by thick purple line.
The lower and upper BKT transition points are $T_{c2}=0.09(2) J_{zz}$ and
$T_{c1}=0.035(15) J_{zz}$~\cite{TFIM_MonteCarlo}. The green solid-dot line refers
to first-order transition while the upper and lower red dots at $B=0$
axis correspond to BKT and inverted BKT transitions, respectively.
The phase boundaries (dash lines), when $B$ is very small, are 
difficult to be captured in the current algorithm and are schematic here. 
The QMC simulation is performed with system sizes $L=6,12,24$ 
with the periodic boundary condition.}
\label{Fig:phases_BT} 
\end{figure}

The thermal melting of the three-sublattice order takes a two-step
manner~\cite{PhysRevB.97.085114,Damle2015,Moessner2000,TFIM_MonteCarlo}
and is also clearly identified in the order parameter histogram as
is shown in Fig.~\ref{Fig:histogram}. At the low-temperature phase
${T<T_{c1}}$ that is proximate to the ground state, the $\mathbb{Z}_{6}$
clock term is relevant such that the phase of $\phi$ is pinned to
six equivalent angles, and we have the three-sublattice long-range
ordered state. This can be seen in the angular histogram plot of the
order parameter $\psi$ in Fig.~\ref{Fig:histogram}(a) where the
six-fold variation is shown. The dual phase at ${T>T_{c2}}$ is the
high-temperature disordered phase where the vortices proliferate.
The higher temperature transition at $T_{c2}$ belongs to the BKT
universality class, while the lower temperature transition at $T_{c1}$
is dual to the high temperature one and hence is called the ``inverse
BKT'' transition. Unlike the 2d XY model with a global U(1) symmetry
where $T_{c1}$ and $T_{c2}$ coincide, in our case $T_{c1}$ and
$T_{c2}$ do not coincide due to the presence of ${\mathbb{Z}}_{6}$
clock term in the free energy of Eq.~\eqref{eq:EFT}. In the intermediate
temperature ${T_{c1}<T<T_{c2}}$ we have an extended phase where both
vortices and the clock anisotropy become irrelevant. The irrelevance
of clock anisotropy indicates an emergent continuous U(1) symmetry
that is shown in Fig.~\ref{Fig:histogram}(b). Due to the emergent
U(1) symmetry, the system behaves just like the low-temperature quasi-long-range
ordered phase of the XY model without any anisotropy term and supports
an algebraic spin correlation, and this thermal regime with ${T_{c1}<T<T_{c2}}$
is referred to as BKT 
phase~\cite{PhysRevB.97.085114,Damle2015,Moessner2000}. 
As long as the ground state is in the three-sublattice ordered phase,
this BKT phase generically occurs in the finite temperature regime
{\sl regardless of the parameters}. For this reason, we plot the finite
temperature phase diagram in Fig.~\ref{Fig:phases_BT} with a single
choice of transverse field ${h/J\approx0.65}$ 
that might be appropriate for TmMgGaO$_4$ 
inside the three-sublattice ordered phase.

\begin{figure}
\includegraphics[width=1\columnwidth]{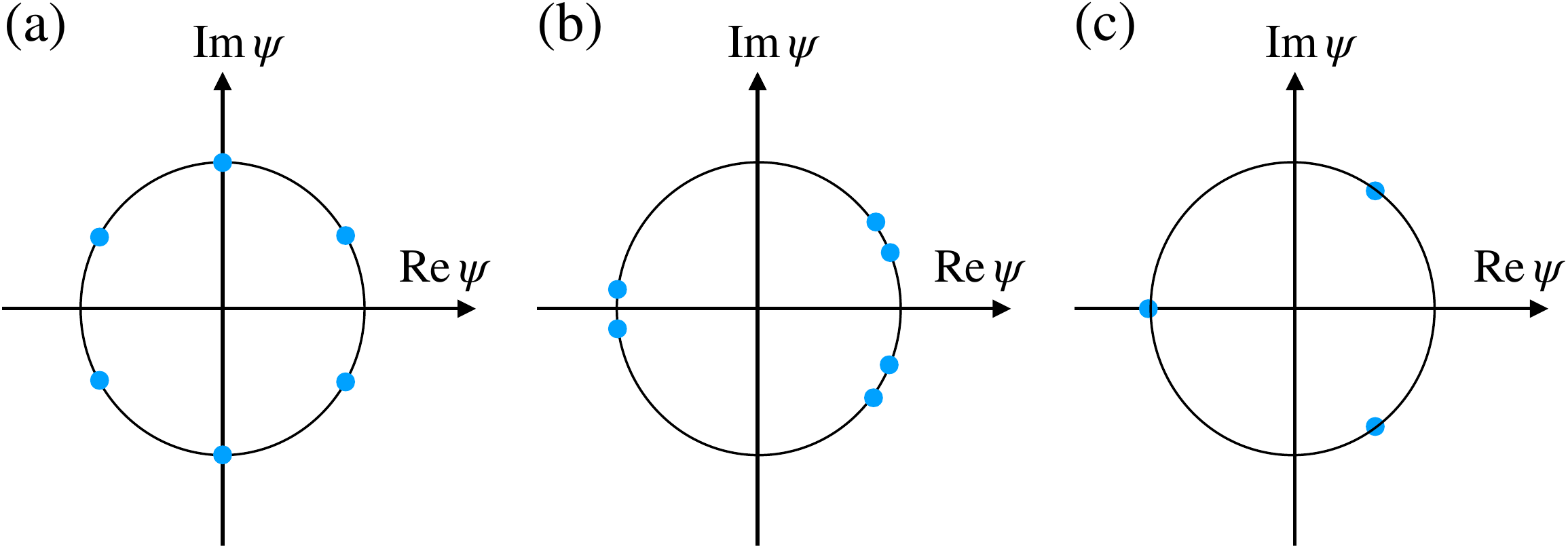}
\caption{Evolution of the order parameter manifold (marked as blue dots) with
external magnetic field $B$. (a): the ``I'' state with \textbf{$B=0$;}
(b) the ``I'' state with small \textbf{$B\protect\neq0$;} (c):
the intermediate three-sublattice ``II'' state with larger $B$.
The the symmetry of order parameter manifolds are $\mathbb{Z}_{6}$,
$\mathbb{Z}_{3}\times\mathbb{Z}_{2}$ and $\mathbb{Z}_{3}$ for three
cases, respectively.}
\label{Fig:OP_B}
\end{figure}

The underlying reason for the finite-temperature BKT physics in this
context arises from the emergent U(1) symmetry. This emergent U(1)
symmetry, however, no longer holds in the presence of external magnetic
fields. The magnetic field breaks the time-reversal symmetry and brings
about a $\mathbb{Z}_{3}$ clock anisotropy to the system~\cite{Damle2015},
\begin{equation}
H_{3}=v_{3}(\psi{}^{3}+\psi^{*}{}^{3}),
\label{eq:H3}
\end{equation}
with $v_{3}$ linearly proportional to $B$. This $\mathbb{Z}_{3}$
clock term is always relevant at the phase transition. Therefore,
the successive BKT transition scenario in thermal melting as well
as an emergent continuous symmetry are no longer presented. Moreover,
from Eq.~\eqref{eq:EFT} and Eq.~\eqref{eq:H3} we obtain the order 
parameter symmetry for each phases, as is shown in Fig.~\ref{Fig:OP_B}. 
We find that with magnetic field the order parameter symmetry of the
three-sublattice ``I'' is reduced from $\mathbb{Z}_{6}$ 
to $\mathbb{Z}_{3}\times\mathbb{Z}_{2}$.
The symmetry is further reduced to $\mathbb{Z}_{3}$ in the intermediate
three-sublattice ``II'' state.

The finite-temperature phase diagram of the three-sublattice state
is shown in Fig.~\ref{Fig:phases_BT}. According to Fig.~\ref{Fig:phases_BT},
if one lowers the temperature from the trivial high-temperature paramagnetic
phase at ${B=0}$, one experiences two successive transitions at $T_{c2}$
and $T_{c1}$. For BKT transitions the correlation length diverges
too fast near the thermal transition, the diverging behavior of the
specific heat near the transition temperatures cannot be very well
observed experimentally or even numerically. This seems to be what
happens for TmMgGaO$_{4}$: no diverging behavior is revealed in the
specific heat data, instead only tiny anomaly with a slightly broad
peak is shown at ${\sim1}$K~\cite{YueshengTMGO,shen2018hidden}.

With magnetic field in Fig.~\ref{Fig:phases_BT}, the $\mathbb{Z}_{3}$
clock anisotropy is introduced and the BKT scenario breaks down. For
the intermediate ``II'' state that breaks $\mathbb{Z}_{3}$ symmetry,
there is only the first order thermal transition. For the ``I''
state that breaks $\mathbb{Z}_{2}$ symmetry in addition to $\mathbb{Z}_{3}$,
the $\mathbb{Z}_{2}$ and $\mathbb{Z}_{3}$ symmetries should break
at different temperatures, therefore one expects another Ising transition
in addition to the first order transition in the thermal melting.  
Unlike the BKT transitions that are weak and unclear in the heat capacity,
these transitions are expected to show diverging signals (the Ising
transition) or discontinuous signals (the first order transition)
in the thermodynamic measurements, as is shown in the magnetic specific
heat data in Ref.~\onlinecite{YueshengTMGO}. However, for the transitions
involving the ``I'' state where the magnetic field is weak, the
divergent behavior is too weak to be observed experimentally or even
numerically, as the system is close to the $B=0$ point where the
BKT scenario happens.

\subsection{Some experimental implications on BKT physics}

Experimentally, it is typically hard to detect BKT transitions in
magnetic systems. Here we discuss how to determine BKT transition
temperatures from experiments. 
Inside the BKT phase between $T_{c2}$ and $T_{c1}$, the algebraic
spin correlation would lead to quasi-Bragg peak at the wavevector
$K$. In principle, Bragg peaks may be distinguished from quasi-Bragg
peaks by the elastic peak profile at the $K$ point, but this is again
difficult. Further 
neutron scattering studies might be useful to sort out the lower transition 
temperature $T_{c1}$. 

A relevant experimental prediction given by K. Damle in Ref.~\onlinecite{Damle2015}
is the singular uniform magnetic susceptibility along $z$ direction
in part of the BKT phase regime, despite absence of ferromagnetic
order in this system. Due to the small energy scale of the interaction,
the direct susceptibility measurements may not be able to give clear 
signals especially because the impurity and disorder effects
could affect very low-temperature thermodynamic behaviors. 
Somewhat equivalently, it is more convenient for us to examine the $S^z$-$S^z$ 
correlation at the $\Gamma$ point in the neutron scattering measurement.
From the available neutron data for TmMgGaO$_4$~\cite{shen2018hidden},
we observe a clear upturn of the Bragg peak intensity below $\sim1$~K.
This $\Gamma$ point upturn at $\sim1$K may be interpreted as the onset of 
the BKT phase if this upturn is not due to any other reason. 
Thus we identify $T_{c2}$ as $\sim1$K, that is consistent with 
the $T_{c2}$ obtained from the magnetic specific heat. 
From the phase diagram by Isakov and Moessner
in Ref.~\onlinecite{TFIM_MonteCarlo} that indicated the 
phase boundary of the BKT phase, we conclude that $T_{c1}$ 
is $\sim0.5$K. Thus, we postulate that the range of BKT phase 
is from $\sim0.5$K to $\sim1$K for TmMgGaO$_4$.

\begin{figure}
\includegraphics[width=1\columnwidth]{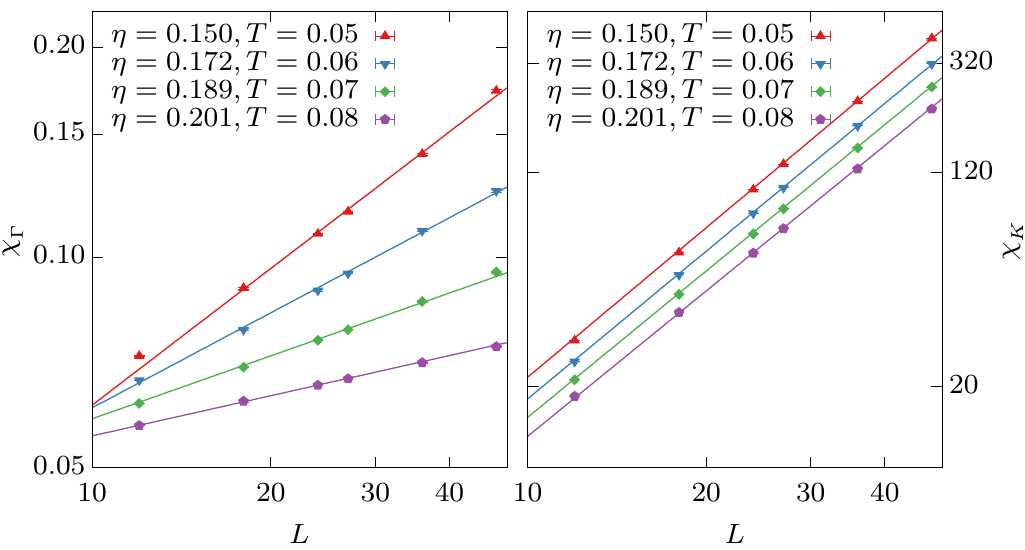}
\caption{Susceptibilities $\chi_{_{\Gamma}}$ and $\chi_{_{K}}$ versus the linear
system size $L$ for four different temperatures when $h=0.65$ in
BKT phase. Solid lines are power-law forms $\sim L^{2-9\eta}$ (on
the left panel) and $\sim L^{2-\eta}$ (on the right panel). Small
systems maybe will deviate from power-law forms due to the effect
of large-finite size scaling.}
\label{Fig:chi}
\end{figure}

Here we numerically examine the power-law behaviors of 
the $S^z$-$S^z$ correlation at $\Gamma$ point and at
$K$ point inside the BKT phase. We have
\begin{equation}
\chi_{_{\Gamma}}=\frac{L^{2}}{\beta}
\Big\langle\Big|\int_{0}^{\beta}d\tau\ m_{_{\Gamma}}(\tau)\Big|^{2}\Big\rangle, 
\label{eq:chi0}
\end{equation}
and the order parameter susceptibility
\begin{equation}
\chi_{_{K}}=\frac{L^{2}}{\beta}
\Big\langle\Big|\int_{0}^{\beta}d\tau\ m_{_{K}}(\tau)\Big|^{2}\Big\rangle ,
\label{eq:chiK}
\end{equation}
where 
${m_{\bf q}\equiv \frac{1}{N} \sum_{i} S_{i}^{z} \exp (i {\bf q} \cdot {\bf r}_i)}$
with $N$ the system size.
According to the previous field theoretical analysis by K. Damle~\cite{Damle2015}, 
they are expected to scale with the system size $L$ as
\begin{eqnarray}
&& \chi_{_{\Gamma}}\sim L^{2-9\eta(T)} ,
\label{eq:chi0sl}\\
&& \chi_{_{K}}\sim L^{2-\eta(T)} ,
\label{eq:chiKsl}
\end{eqnarray}
with $ {1/9 \leq \eta(T) \leq 2/9}$ in part of the BKT phase~\cite{Damle2015}. 
The QMC results are shown in Fig.~\ref{Fig:chi} and
both fit rather well to Eq.~\eqref{eq:chi0sl} and Eq.~\eqref{eq:chiKsl}.
Although there is always no ferromagnetic long-range order in the BKT phase, 
$\chi_{_{\Gamma}}$ is still divergent with the system 
size in part of the phase region.

\section{Dynamic properties from orthogonal operator approach and selective measurements}
\label{sec6}

The previous section deals with the phase diagram and the magnetic
ordering structures. These properties are static magnetic properties.
To provide more information about the system, we here explore the
dynamic properties from the orthogonal operator approach and the selective
measurements. In Sec.~\ref{sec6a}, we explain the ``orthogonal
operator approach''. In Sec.~\ref{sec6b}, we turn to the selective
measurement that directly applies the ``orthogonal operator approach''
for the Tm-based triangular lattice antiferromagnets.

\subsection{Orthogonal operator approach}

\label{sec6a}

Even though the in-plane components are non-zero, they are not visible
from the experiments. These ``hidden order''-like features can be
revealed from an approach called ``orthogonal operator approach''~\cite{GCnonK,GCoctu}.
The notion of ``hidden order'' was introduced into condensed matter
physics in the study of the compound URu$_{2}$Si$_{2}$~\cite{2014Mydosh}.
The order parameter associated with the hidden order does not couple
strongly with the conventional experimental probe such that the order
does not explicitly show up in the usual experimental probes. To identify
the nature of the hidden order, our simple suggestion was to find
the physical observables whose operators do not commute with the proposed
hidden order operators, and at the same time make sure these observables
are ready to detect experimentally. These operators are referred as
``orthogonal operators''. The dynamic correlations or spectra of
these operators would reveal the structure and the nature of the underlying
hidden orders. These thoughts have been explored for the quadrupolar
orders and the octupolar orders of triangular lattice magnets~\cite{GCnonK,GCoctu}
as well as the spin nematics in frustrated magnets~\cite{PhysRevB.88.184430}.

Because the non-vanishing in-plane components are induced by the intrinsic
transverse field, strictly speaking, they are not the Landau symmetry
breaking orders. Nevertheless, their presence and behavior are very
much similar to the roles of the hidden orders and thus can be understood
in a similar manner.

\subsection{Selective measurements}

\label{sec6b}

\begin{figure*}[!t]
\includegraphics[width=0.3\textwidth]{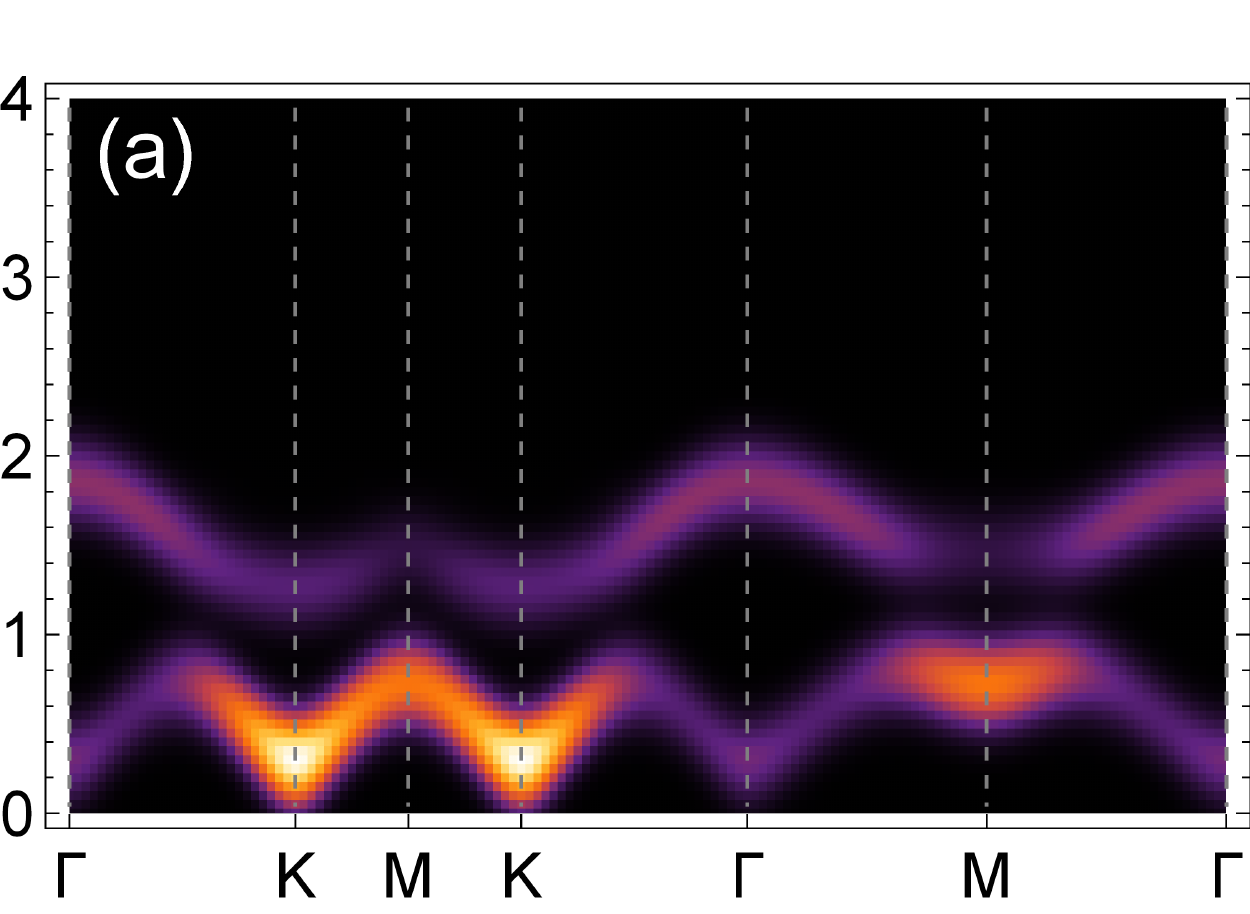}\enspace{} \includegraphics[width=0.3\textwidth]{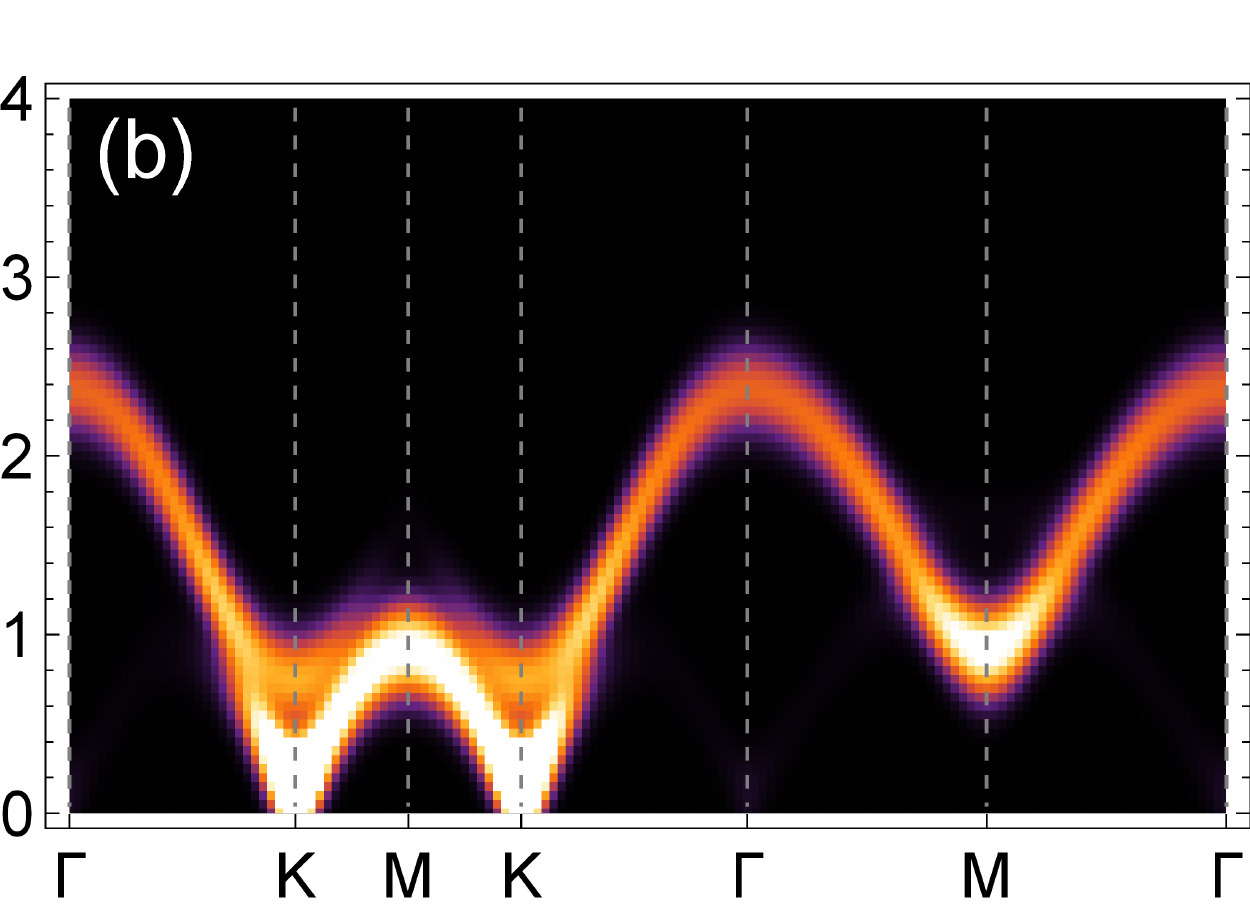}\enspace{}
\includegraphics[width=0.3\textwidth]{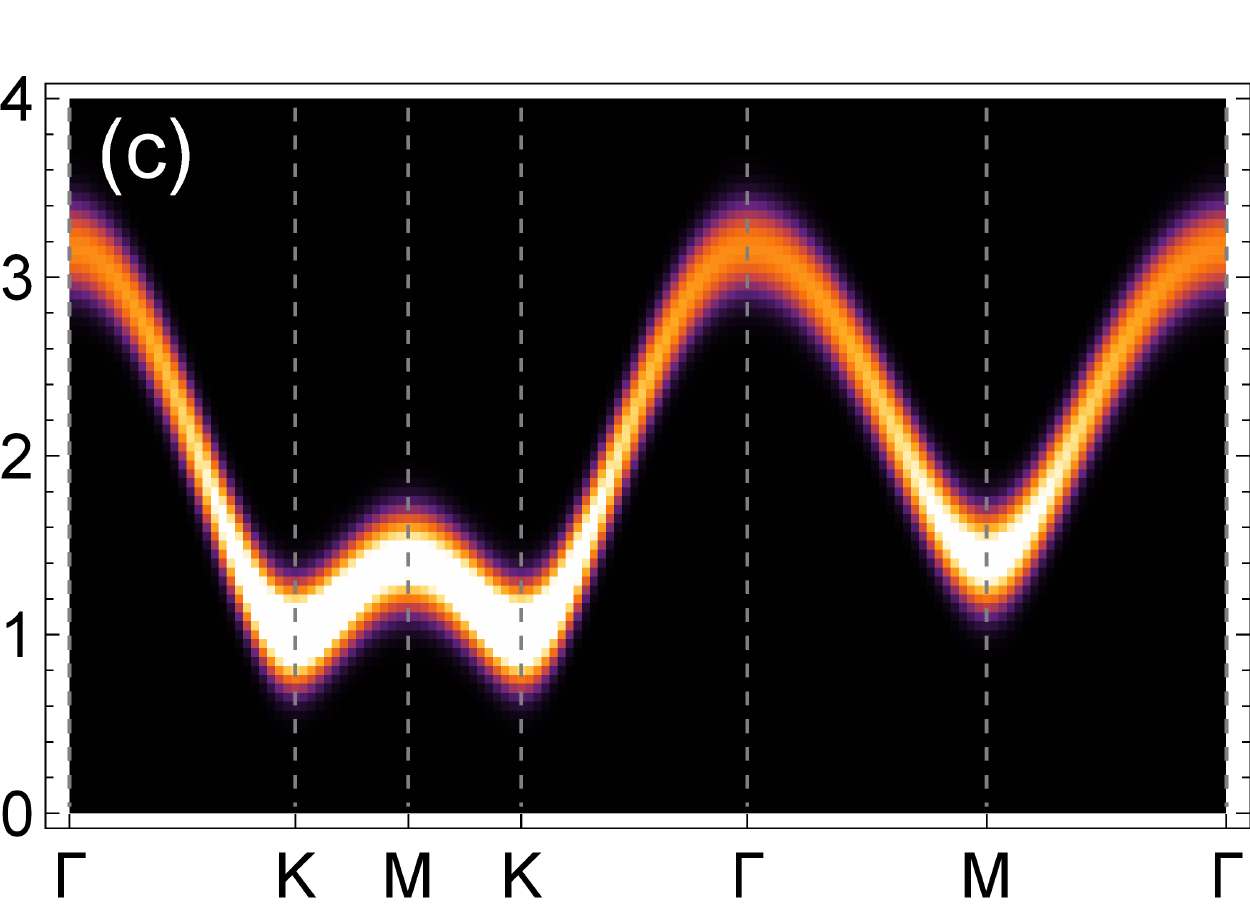} 

\includegraphics[width=0.3\textwidth]{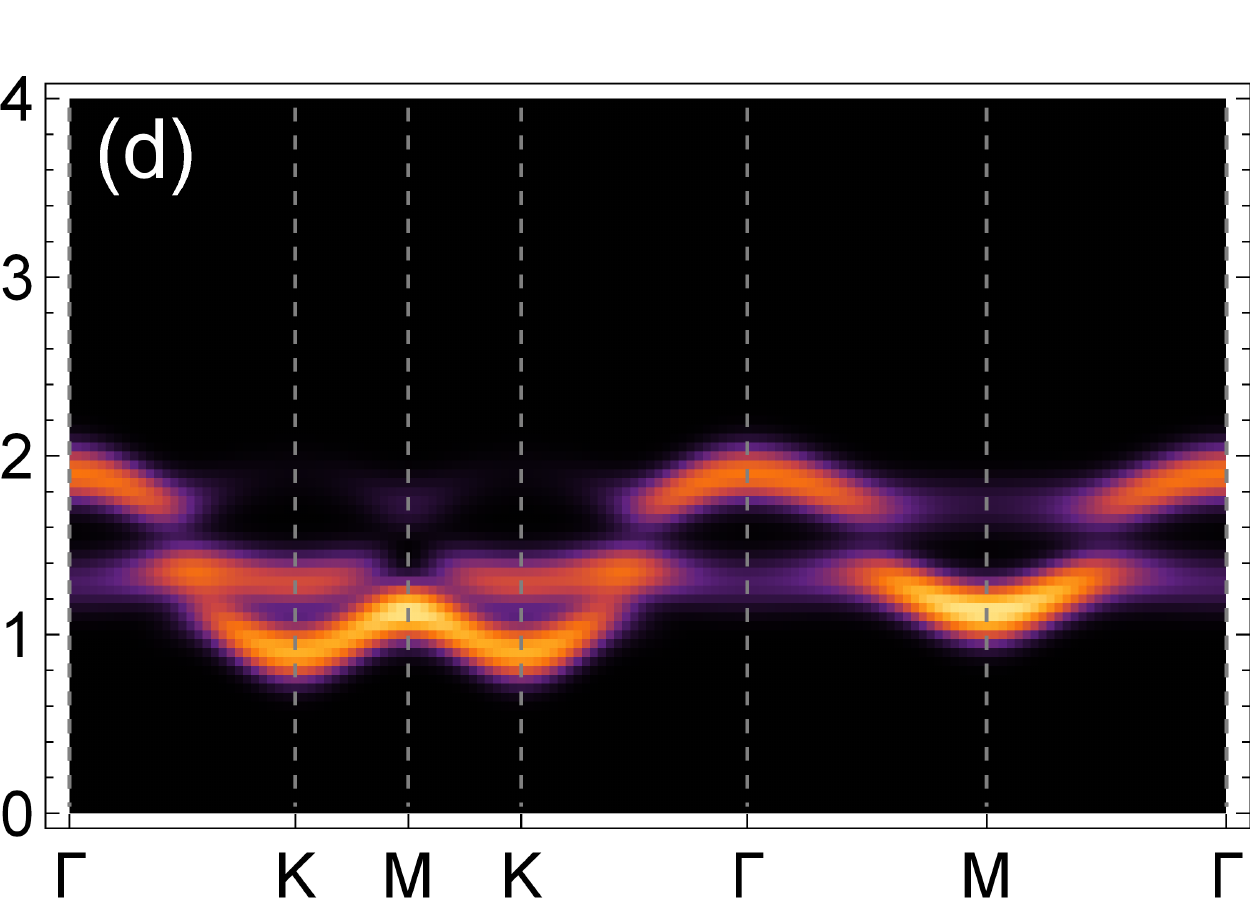}\enspace{} \includegraphics[width=0.3\textwidth]{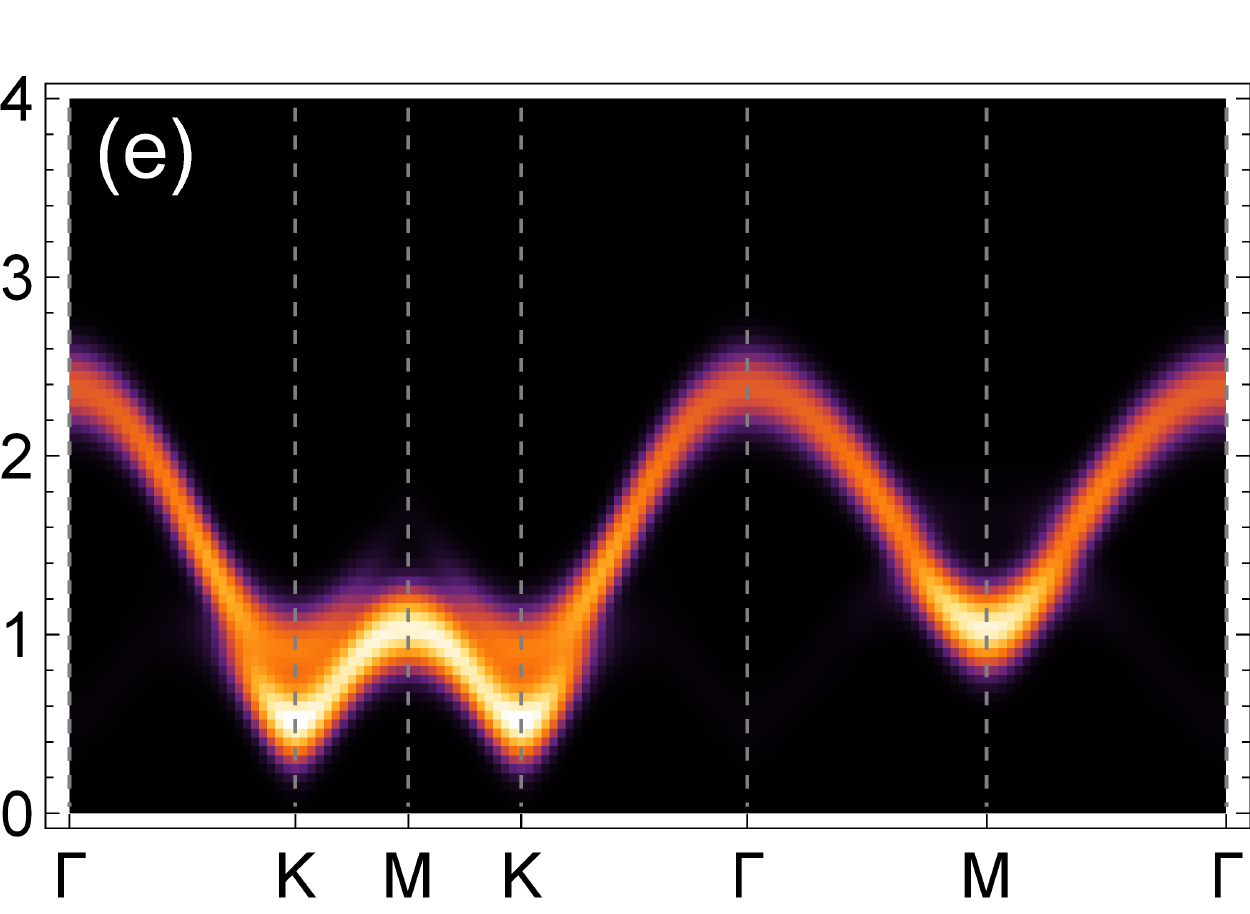}\enspace{}
\includegraphics[width=0.3\textwidth]{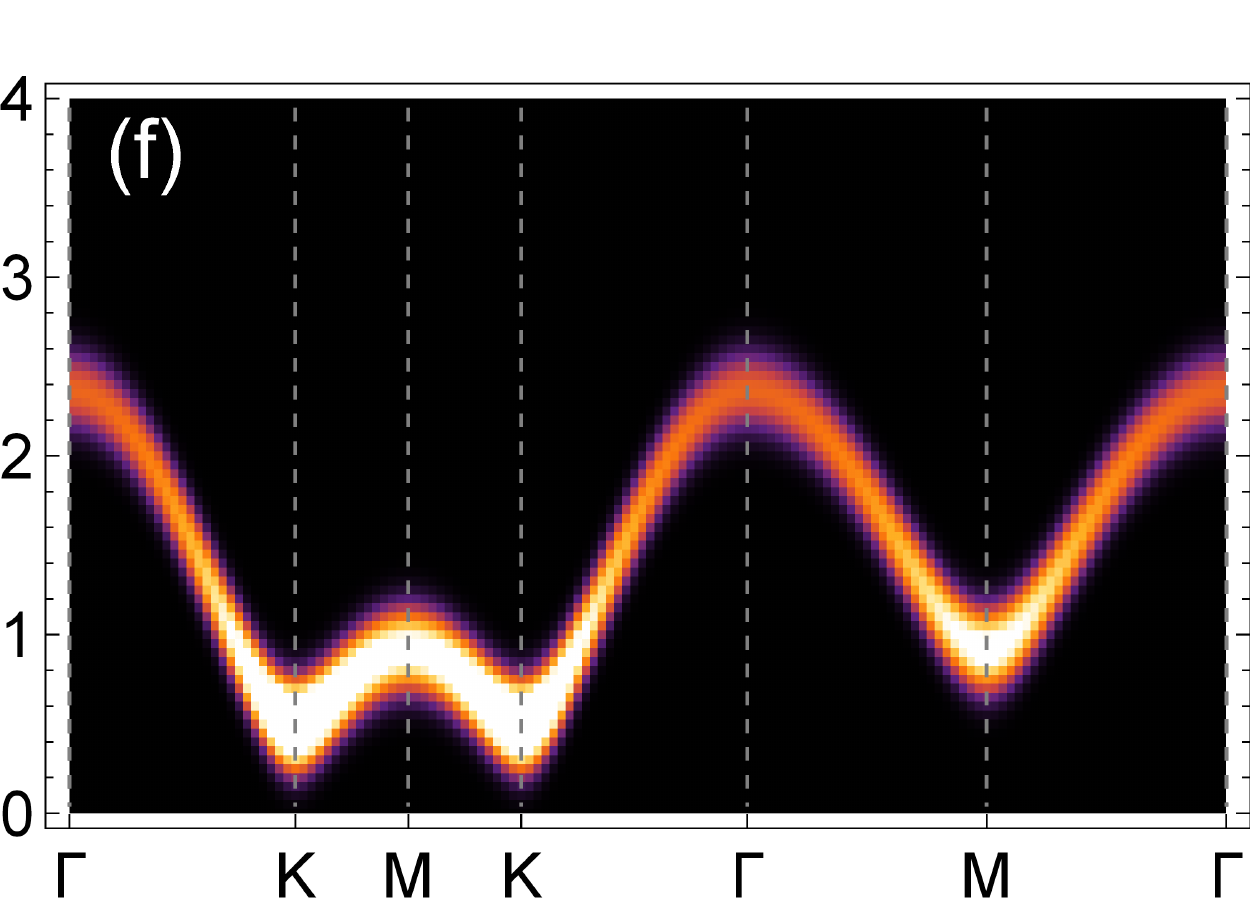} 

\caption{Dynamical correlation function $\mathcal{S}^{zz}(\boldsymbol{q},\omega)$
calculated within the linear spin-wave theory of (a),(b):the three-sublattice
``I'' state, (c),(d): the intermediate three-sublattice ``II''
state and (c),(f): the paramagnetic (or polarized) state. The parameter
we take for the representative points are (a): ${h=0.8}$, ${B=0}$;
(b): ${h=1.3}$, ${B=0}$; (c): ${h=2}$, ${B=0}$, (d): ${h=0.8}$,
${B=1.3}$; (e): ${h=1.3}$, ${B=1.3}$; (f): ${h=1.3}$, ${B=1.8}$.
In all cases we take ${J_{zz}=1}$.}
\label{Fig:excitations} 
\end{figure*}

Having figured out the phase diagrams in the previous section, we
here explain the experimental consequences for the dynamics from the
selective measurements and the orthogonal operator approach. The three-sublattice
order that we found from the model is characterized by the order parameter
$\psi$ defined by the dipolar magnetization, which is directly reflected
as the magnetic Bragg peaks at the $K$ point. Meanwhile, due to the
intrinsic crystal field, there is always a non-vanishing expectation
value in the transverse components that arises not from the spontaneous
symmetry breaking but from the intrinsic polarization effect. Since
the transverse components are the magnetic multipoles, they do not
directly couple to the neutron spins hence are hidden in the neutron
probes. Due to the peculiar local moment structure of this system,
however, the elementary excitations of the multipole moment can be
measured in the dynamic probes such as the inelastic neutron scattering,
owing to the non-commutative relation between the dipole and the multipole
moments. This specific idea was initially pointed out in the context
of the non-Kramers doublets~\cite{GCnonK} and also applies here.
As the neutron spins only directly couple to the dipole components,
in the inelastic neutron scattering what is measured is the $S^{z}$-$S^{z}$
correlation 
\begin{equation}
{\mathcal{S}}^{zz}(\boldsymbol{q},\omega)
=\frac{1}{2\pi N}\sum_{ij}\int_{-\infty}^{+\infty}\mbox{d}t\,
  e^{i\boldsymbol{q}\cdot(\mathbf{r}_{i}-\mathbf{r}_{j})-i\omega t}
  \langle S_{i}^{z}(0)S_{j}^{z}(t)\rangle,
\label{eq3}
\end{equation}
and the transverse component correlation is not directly visible in
the neutron scattering measurement. Based on the above selective measurement,
a regular neutron scattering measurement would behave like a polarized
neutron measurement that automatically selects the $S^{z}$-$S^{z}$
correlation. As the neutron spin detects the longitudinal dipole moments,
it ``flips'' the multipole moment that is orthogonal to the dipole
moment, creating the coherent spin-wave excitations. Therefore, in
an inelastic neutron scattering experiment, what is measured is the
elementary excitation of the multipole components that contains the
information on the underlying hidden multipole structures. We have
calculated the dynamic structure factors for three representative
parameters. The results are shown in Fig.~\ref{Fig:excitations}.
For the paramagnetic (or Ising disordered) side, there is only one
branch of excitation, reflecting the uniform structure of the paramagnetic
(or Ising-disordered) phase with a ``ferro-multipole order '' $\langle S^{y}\rangle$
(see Fig.~\ref{Fig:excitations}(c)). If another Tm-based 
triangular lattice material is located in
this parameter regime and phase, there will be no transition through
all temperatures but the excitation spectrum surprisingly becomes more coherent
as the temperature is lowered despite the absence of any ordering. 
This phenomenon can be quite striking from the experimental perspective.

\begin{figure*}[t!]
\includegraphics[width=0.9\textwidth]{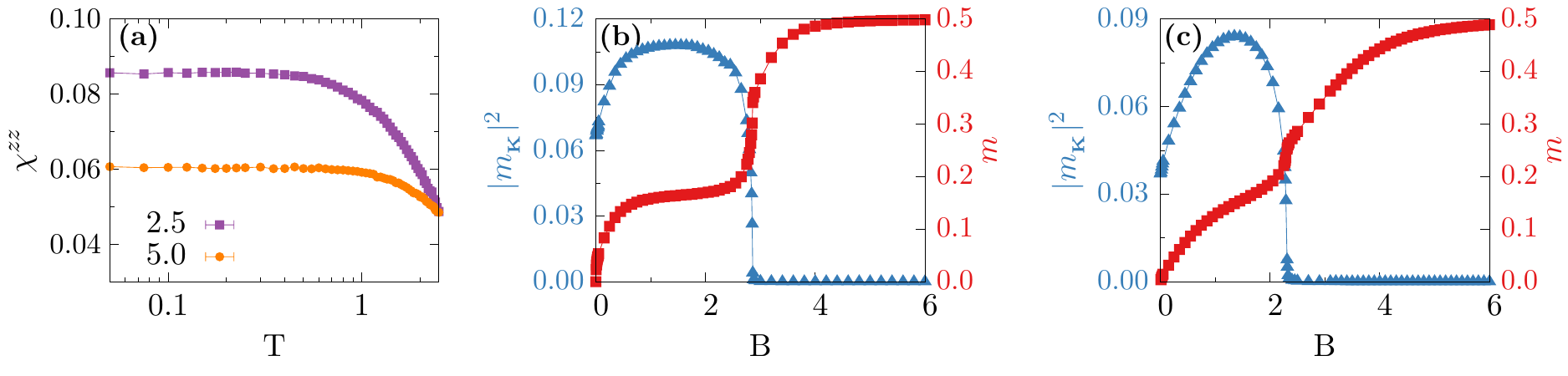} 
\caption{(a) The magnetic susceptibility $\chi^{zz}$ versus the temperature
$T$ for the paramagnetic (or the Ising disordered) state. Here we
take the parameter ${J_{zz}=1}$, ${h=2.5,5.0}$. The magnetic susceptibility
is defined as ${\chi^{zz}={\partial m}/{\partial B}}$ where ${m=(\sum_{i}S_{i}^{z})/N}$
is the dipolar magnetization per site. (b) and (c): Magnetization
$m$ (in red) and  the magnetic Bragg peak $|m_{_{K}}|^2$
(in blue) for the three-sublattice state at low temperatures. The
parameter we take is ${J_{zz}=1}$ and for (b): ${h=0.25}$, for (c):
${h=0.65}$. All results are calculated through QMC with the system
size $L=12$ and ${\beta=80}$. The error bars are much smaller than
the points.}
\label{Fig:thermodynamics} 
\end{figure*}

Meanwhile, for the three-sublattice ordered state, one can roughly
identify two branches of excitations in Fig.~\ref{Fig:excitations}(b)
and clearly identify two branches of excitations in Fig.~\ref{Fig:excitations}(a).
The experimental situation~\cite{shen2018hidden} in TmMgGaO$_{4}$
is more close to Fig.~\ref{Fig:excitations}(b) that shows a reasonable
agreement with the experimental data. In Fig.~\ref{Fig:excitations}(b),
we choose the specific parameter ${h/h_{c}^{\text{MF}}\simeq0.87}$
where $h_{c}^{\text{MF}}$ is the critical field of the mean-field
theory. The counting of the branch number immediately brings up a
question that the number of branches in the experiment is inconsistent
with the number of magnetic sublattices. This question was \textsl{not}
raised in Ref.~\onlinecite{shen2018hidden} and is addressed here.
In fact, our honest linear spin-wave calculation of the full spectra
in the Appendix.~\ref{append} gives three branches of dispersions
that correspond to the three-sublattice magnetic structure. The reason
that the $S^{z}$-$S^{z}$ correlation looks like two branches is
because the selective measurement makes the intensity of part of the
excitation spectra rather weak such that the spectra look like two
branches. This indicates the \textsl{incompleteness} of the selective
measurements. Other dynamic measurements such as optics and THz may
avoid the selective measurement issue and provide complementary information
about the excitations here.

For the specific TmMgGaO$_{4}$ material, the previous neutron scattering
experiment shows a tiny spin gap at the $K$ point~\cite{shen2018hidden}.
This is expected as the model and the system do not have any continuous
symmetry to support any gapless Goldstone mode. The reason that the
gap is tiny is a common consequence of the quantum order by disorder~\cite{PhysRevLett.109.167201}
for the TFIM with the antiferromagnetic Ising interaction on the triangular
lattice.
This tiny-gapped mode is sometimes referred as pseudo-Goldstone mode
as it appears as a breaking of continuous symmetry at the quadratic 
or linear spin-wave theory level~\cite{PhysRevLett.109.167201}.

\section{More effects from external magnetic field}

\label{sec7}

The external magnetic field not only enriches the phase diagram but
also generates more experimental consequences to be examined. In this
section, we will first focus on the static properties of the system
such as the magnetization and static spin structure factor under the
external magnetic field, and then explore the dynamic properties of
the system.

\subsection{Susceptibility, magnetization, and non-monotonic ordering}
\label{sec7a}

Except in a small parameter regime near the critical point (see Fig.~\ref{Fig:phases})
where the magnetic field could drive a re-entrant transition by crossing
the three-sublattice ordered one, we do not expect phase transitions
upon increasing temperatures with or without the external field for
the paramagnetic (or the Ising disordered) state that preserves all
the lattice symmetries. This behavior is fundamentally different from
those of the three-sublattice ordered state, which can be used to
identify these two phases without performing the neutron scattering
experiments. The magnetic susceptibility $\chi^{zz}$ as a function
of the temperature is calculated via QMC and is plotted in Fig.~\ref{Fig:thermodynamics}(a).
At high temperatures the magnetic susceptibility satisfies the Curie-Weiss
law with $\chi^{zz}\simeq{C}/{(T-\Theta_{\text{CW}})}$, where $C$
is the Curie constant and ${\Theta_{\text{CW}}=-{3}J_{zz}/2}$ where
one can extract the exchange parameter $J_{zz}$. There is a crossover
to the low temperature behavior where $\chi^{zz}$ saturates to a
constant. This is because the Hamiltonian does not have any continuous
symmetry and the total magnetization is not a good quantum number
to label the many-body states. Within a simple mean-field theory,
we find the low-temperature $\chi^{zz}$ is given as 
\begin{equation}
{\chi^{zz}|}_{T\rightarrow0}\approx\frac{1}{6J_{zz}+2h}.
\label{eq:lowT}
\end{equation}
Compared to the QMC data, we find that at large $h$, they coincide
very well. Therefore, one can extract the model parameter $J_{zz}$
and $h$ simply from the high-temperature and low-temperature behaviors
of $\chi^{zz}$ if the system is located in the paramagnetic phase.
The above relation is especially useful if the ground state of the 
system is in the Ising disordered phase. 

We continue to discuss the magnetization process of the three-sublattice
state that can be relevant for the specific material TmMgGaO$_{4}$. 
In the absence of the external magnetic field, the three-sublattice ordering 
arises from the quantum order-by-disorder mechanism. The spin excitation gap 
is relatively small (see Fig.~\ref{Fig:excitations}(a),(b)). This property
makes the three-sublattice state fragile against the external magnetic
field. A small external field at $B_{c1}$ will cause the closing
of the spin gap and drive the system towards the intermediate quasi-plateau
state. Here ``quasi-plateau'' is used as the total magnetization
is not conserved. With the increasing external magnetic field, the
spin gap re-opens and the intermediate state becomes stable. Further
increasing magnetic field the spin gap drops until the system is driven
to a polarized state by the magnetic field at $B_{c2}$ above which
the system is smoothly connected to the fully polarized one.

The presence of the intermediate quasi-plateau state renders the the
magnetization process non-trivial, as it is shown in Fig.~\ref{Fig:thermodynamics}
from the QMC calculation. For a small $h$, the magnetization curve
shows a clear $1/3$ quasi-plateau feature in the intermediate regime.
Meanwhile, deep in the quasi-plateau state, the system has an approximate
``2-up-1-down'' structure that contributes to a robust three-sublattice
ordering compared to the case without the external field. Therefore,
in the elastic neutron experiments the intensity of the magnetic Bragg
peak $|m_{_{K}}|^2$ (proportional to $|\psi|^{2}$) is expected
to show \textsl{non-monotonic} behaviors: deep in the quasi-plateau
state the intensity is large, while approaching the three-sublattice
state I the intensity is expected to decrease; the intensity is also
expected to decrease when the field is large enough where the system
becomes nearly polarized (see Fig.~\ref{Fig:thermodynamics}(b)).

For the case relevant with TmMgGaO$_{4}$ where the transverse field
$h$ is comparable to the exchange $J_{zz}$, in the quasi-plateau regime
the ``2-up-1-down'' structure is heavily distorted, therefore the
quasi-plateau feature of the intermediate regime is not clearly observed
in the magnetization curve. Instead, the line shape curves slightly
downwards at $B_{c2}$ (see Fig.~\ref{Fig:thermodynamics}(c)). This
feature is found in the magnetization data of TmMgGaO$_{4}$ at about
2.5T, which marks the transition field 
$B_{c2}$~\cite{Cevallos2018,YueshengTMGO,shen2018hidden,2019arXiv191202344L}.
Above $B_{c2}$ the system becomes polarized, but not fully aligned
along the $z$ direction due to the presence of the intrinsic transverse
field. In order to allow the magnetization approach the saturation
value, a larger external field has to be applied. This feature is
in a stark contrast to ordinary systems where the internal transverse
field is absent. For the magnetic Bragg peak $|m{_{K}}|^2$,
the {\sl non-monotonic} behavior persists with large transverse fields (see
Fig.~\ref{Fig:thermodynamics}(c)).

\subsection{Dynamic properties in magnetic fields}

Here we discuss the dynamical properties in presence of external magnetic
field. With applying small external magnetic field, the gap first
decreases and closes at $B_{c1}$. As $B_{c1}$ is typically small,
this phenomenon is subtle and can be hard to be observed experimentally.
With increasing magnetic field the system gap reopens across $B_{c1}$
as it enters the intermediate ``II'' regime. In the ``II'' regime
the gap behaves non-monotonically: the gap first increases, reaches
maximum and drops, until the system becomes polarized at $B_{c2}$
via a first-order transition. As this transition being first-order,
the gap does not close across $B_{c2}$.

We have calculated the spin excitation spectra with magnetic field,
as is shown in Fig.~\ref{Fig:excitations}(d)-(f). From Fig.~\ref{Fig:excitations}(d)
we can clearly see three spin-wave branches, consistent with the three-sublattice
magnetic order. Therefore, the ``selection rule'' breaks down with
magnetic field. Another observation is that there remains non-zero
intensity even in the fully polarized state, see Fig.~\ref{Fig:excitations}(f).
This is a peculiar feature for our non-degenerate dipole-multipole
doublet systems due to the intrinsic transverse field: the spins are
tilted to acquire non-zero transverse components, which results the
non-vanishing intensity in the polarized state.

\section{Discussion}
\label{sec8}

\subsection{Summary for TmMgGaO$_{4}$}

In this paper, we have performed a theoretical study on the triangular
lattice transverse field Ising model relevant with the TmMgGaO$_{4}$
material. We clarify the intrinsic origin of transverse field of this
material as the crystal field splitting. We established the full phase
diagram by combining the mean-field theory and the QMC simulation.
We discuss the continuous symmetry and BKT physics that emerge in the thermal 
melting and at the quantum critical point. We explain the properties of phases
in the neutron scattering measurement and the thermodynamic experiments.
The available experimental data show that this material at zero field
is well consistent with the TRIM with the
three-sublattice intertwined dipolar and multipolar ordered ground state. 

We mention a couple recent works on TmMgGaO$_{4}$ and the transverse
field Ising model on triangular lattice. A recent numerical-oriented
work~\cite{li2019ghost} explored our proposed effective model for
TmMgGaO$_{4}$ using more updated numerical techniques and focused
on the numerical aspects of the model. 
Their results supported
the validity of the TFIM for TmMgGaO$_{4}$. 
Ref.~\onlinecite{li2019ghost} suggested the system first enters the BKT phase 
at $\sim 4$K and then enters the 3-sublattice ordered state at $\sim 1$K. 
This differs from the results of the current work, 
where we have $\sim 1$K for the upper BKT transition 
and $\sim 0.5$K for the lower one. 
They further established the roton mode at the $M$ point inside the three-sublattice
ordered state. This is probably due to the presence of the 
second-neighbor interaction. One may understand this in analogous 
with the supersolidity and the roton mode in the spin-1/2 XXZ model or repulsive 
hard-core boson model at half-filling
on the triangular lattice where the roton condensation leads to the 
$S^z$ order~\cite{PhysRevLett.95.127205,PhysRevLett.95.127207} on top of the 
transverse component order.
The difference is that, the TFIM here develops an emergent U(1) 
symmetry and cannot have supersolidity  
while the XXZ model has a global U(1) symmetry.
One may further consider the effect of the dipole-dipole interaction
 and other effects on this roton mode. 
 
Another quite recent
experimental work~\cite{2019arXiv191202344L} supplemented the early
thermodynamic results~\cite{YueshengTMGO} with neutron diffraction
measurements and corrected the early claim of a pure Ising model with
more analysis. In the new work~\cite{2019arXiv191202344L} the authors
added the transverse field and suggested the Mg/Ga site disorder could
create a (random) distribution of the transverse field. They argued
that this site disorder could be the origin of their proposed ``partial
up-up-down'' order. Based on the neutron diffraction and thermodynamic
measurements, Ref.~\onlinecite{YueshengTMGO} compared the parameters
of different couplings of the model with finite-size exact diagonalization
calculation and supported the proposal of a TFIM with random disorders.
Since Ref.~\onlinecite{YueshengTMGO} raised the possible issue of
random disorders, we agree that the quantitative behaviors of the
thermodynamic results might be more sensitive to disorder effects
on the exchange and ``$g$'' factors as well as the residual coupling
to the high-order multipolar moments. On the other hand, the random
exchange and/or the random transverse field would lead to the line-broadening
with a similar range of energies in the spin-wave spectrum~\cite{YueshengTMGO}.
Although a well-defined spin-wave spectrum with a clear dispersion
was recorded in the inelastic neutron scattering measurement and reported
in Ref.~\onlinecite{shen2018hidden}, we still think more data-rich
experiments are needed at this stage if one hopes to extract more
quantitive information. For example, one could carry out the inelastic
neutron scattering measurements as a function of the external magnetic
field and establish the excitation spectrum in the (more robust) three-sublattice
ordered state of the phase II. One can combine the results of zero
field and finite fields and give an estimate of the transverse field
distribution from the linewidth of the excitations after subtracting
the broadening due to the magnon interactions and the instrument resolution.

In our analysis here, we did not consider the random disorder effect
that was actually raised after the first online version of the
current work in Ref.~\onlinecite{YueshengTMGO}, and were unable
to provide or address much more detailed numerical and quantitative aspects
that relate to the experiments quantitatively. We focus more on the generic
and qualitative physics that may be more robust in the experiments.
Regardless of the specific material,
it will be interesting to understand the fate of the finite-temperature 
BKT phase in the presence of quenched random disorders,
and this may be analyzed with the perturbative renormalization calculation
within the BKT phase.

\subsection{Connection to upper branch magnetism}

In fact, the magnetism of TmMgGaO$_{4}$ belongs to the category of
systems with ``upper branch magnetism''. The notion of ``upper
branch magnetism'' was introduced in Ref.~\onlinecite{Liu2019}.
It refers to the case where the local crystal field environment simply
favors a non-magnetic state while the superexchange interaction prefers
magnetic states of some sort. For the specific illustrative example
in Ref.~\onlinecite{Liu2019}, the local crystal field ground state
is a singlet and the first excited states are a two-fold degenerate
doublet. The specific system over there was modelled as an effective
spin-1 magnet, and the crystal field splitting was modelled as a single-ion
anisotropy for the spin-1 moment.

In what sense is TmMgGaO$_{4}$ regarded as ``upper branch magnetism''?
The magnetism cannot occur if there is no exchange interaction between
the Tm local moments. Crudely speaking, it is the exchange interaction
that ``drag down'' the excited energy level. More precisely, it
is the non-trivial quantum mechanical interplay between the intrinsic
transverse field and the Ising exchange that gives rise to the magnetism
and the associated coherent excitation. What do we expect experimentally
if the system is controlled more by the single-ion physics? The magnetism
will be gone. Despite that, the coherent magnetic excitation would
persist. This may occur in some other systems.

\subsection{Extension to other Tm-based compounds}

A series of rare-earth triangular lattice magnets has been summarized
in Ref.~\onlinecite{GCnonK}. We expect that other materials, especially
some Tm-based compounds, can be also described by the TFIM, and share
similar physics with TmMgGaO$_{4}$. The Tm-based magnetism is not
a common subject in quantum magnetism of the rare-earth systems. Some
of the insights that we learn from TmMgGaO$_{4}$ could be applied
to other Tm-magnets. In the following, we survey the existing Tm-magnets
and explain the physics in them.

\subsubsection{Tm spinels and Tm pyrochlores}

The Tm spinel, MgTm$_{2}$Se$_{4}$, has been studied by the neutron
scattering measurement~\cite{MgTm2Se4}. The crystal electric field
states were carefully studied in Ref.~\onlinecite{MgTm2Se4}. It
turns out that the crystal field ground state and the first excited
state are similar to the ones in TmMgGaO$_{4}$. They are separated
from other excited levels by an energy gap more than 10meV. The wavefunctions
of the lowest two states are 
\begin{eqnarray}
|\Psi_{g}\rangle & = & 0.66960|6\rangle+0.14821|3\rangle+0.24361|0\rangle\nonumber \\
 &  & \quad-0.14821|{-3}\rangle+0.66960|{-6}\rangle,\\
|\Psi_{e}\rangle & = & -0.70097|6\rangle-0.092966|3\rangle\nonumber \\
 &  & \quad-0.092966|{-3}\rangle+0.70097|{-6}\rangle,
\end{eqnarray}
and the energy separation between them is about 0.885meV. Similar
to TmMgGaO$_{4}$, one could introduce an effective spin-1/2 degree
of freedom that operates on these two states. Here, we propose a relevant
model for MgTm$_{2}$Se$_{4}$ would be a transverse field Ising model
on the pyrochlore lattice, 
\begin{eqnarray}
H_{\text{MgTm\ensuremath{_{2}}Se\ensuremath{_{4}}}}=J_{z}\sum_{\langle ij\rangle}S_{i}^{z}S_{j}^{z}-h\sum_{i}S_{i}^{y}-B\sum_{i}(\hat{n}\cdot\hat{z}_{i})S_{i}^{z},
\end{eqnarray}
where $\hat{z}_{i}$ is the local $[111]$ axis, $h$ ($B$) is the
intrinsic transverse (external magnetic) field, and $\hat{n}$ is
the direction of the external magnetic field. Ref.~\onlinecite{MgTm2Se4}
has suggested a vanishing $g$-factor for the Tm$^{3+}$ ion. Based
on our experience and the magnetization measurement in Ref.~\onlinecite{MgTm2Se4},
however, we think that the field would primarily couple to the local
$z$ component of the local moment. The magnetic moment of MgTm$_{2}$Se$_{4}$
can be read from the saturated magnetization after considering the
fact that the magnetic moment is oriented along the local {[}111{]}
direction of each sublattice. The sign of $J_{z}$ plays an important
role in determining the quantum ground state of MgTm$_{2}$Se$_{4}$.
It is ready to obtain that, the Curie-Weiss temperature has ${\Theta_{CW}=J_{z}/2}$
and does not depend on the direction of the probing field. Unfortunately,
the Curie-Weiss temperature is unknown for this material. If ${J_{z}>0}$,
one could establish a pyrochlore ice U(1) spin liquid ground state
when $h$ is small and develop a quantum transition into a disordered
state when $h$ is large. If ${J_{z}<0}$, then one simply has the
usual phase diagram of the ferromagnetic Ising model.

The Tm-based pyrochlore has rarely been studied. The crystal field
levels of Tm$_{2}$Ti$_{2}$O$_{7}$ were computed in Ref.~\onlinecite{Re227}.
It was found that, the crystal field ground state is a singlet with
the wavefunction, 
\begin{eqnarray}
|\Psi_{g}\rangle=0.147|{6}\rangle-0.692|{3}\rangle-0.692|{-3}\rangle-0.147|{-6}\rangle,
\end{eqnarray}
and the first excited state is a doublet with two-fold degeneracy.
This crystal field level setting is identical to the specific case
that was considered in Ref.~\onlinecite{Liu2019}. The energy separation
between the ground state singlet and the first excited doublet is
of the order of 10meV, so it is not in the weak crystal field regime
for the rare-earth magnets. It is likely that other isostructural
Tm-based pyrochlores could have a smaller crystal field gap and allow
more interesting magnetism to happen.

\subsubsection{Tm honeycomb lattice and Tm Kagom\'{e} lattice magnets}

Here we extend some of our thoughts to other two-dimensional systems.
We start with the honeycomb magnet RNi$_{3}$Al$_{9}$ where R is
the rare-earth ion and Tm is a member of them~\cite{doi:10.1143/JPSJS.80SA.SA080}.
These materials have both conduction electrons and local moments,
so it is a conductor and there is a Kondo physics in some of them.
The local moment magnetism is from the rare-earth moments. We focus
the discussion on the Tm-based materials. Other materials in this
family such as the Yb-based ones could involve Kitaev and other anisotropic
spin interactions between the local moments and also worth a further
investigation. The magnetization measurement in the single crystal
sample of TmNi$_{3}$Al$_{9}$ is quite similar to the one in TmMgGaO$_{4}$,
where the out-of-plane response is dominant and the in-plane response
is negligible. There are two possibilities for the Tm magnetism in
TmNi$_{3}$Al$_{9}$. The first possibility is that the Tm local moment
is a (degenerate) non-Kramers doublet. The other possibility is that
the Tm local moment is a non-degenerate dipole-multipole doublet like
the one in TmMgGaO$_{4}$, and the effective model for the Tm magnetism
would be a transverse field Ising model. Due to the presence of the
itinerant electrons, the Ising interaction may involve further neighbors.
This material develops a magnetic order at $2.9$K from the thermodynamic
and transport measurements. From the experience about TmMgGaO$_{4}$,
we expect a coherent excitation spectrum. This may be confirmed by
further experiments with neutron scattering measurements.

The Tm-based Kagom\'{e} 
magnets have been explored recently~\cite{PhysRevB.98.174404,PhysRevB.95.104439,PhysRevLett.116.157201},
and effective spin-1/2 degrees of freedom are used to describe the
Tm magnetism. Unlike the triangular lattice and the honeycomb lattice,
the point group symmetry does not involve an on-site three-fold rotation,
and there is no non-Kramers doublet on the Kagom\'{e} lattice. Thus there
is always an intrinsic splitting between the two relevant crystal
field levels of the Tm$^{3+}$ ion. Because the Tm$^{3+}$ singlets
are not the same kind of singlets as the ones in TmMgGaO$_{4}$, the
exchange part of the interaction is not simply be the Ising model.

\subsubsection{Tm double perovskites}

Another class of the Tm magnets is the Tm-based double perovskite.
Unlike the rare-earth pyrochlores and the rare-earth triangular lattice
magnets, these materials have not been well studied before. Here Tm
ions form a FCC lattice. Only two Tm-based double perovskites Ba$_{2}$TmSbO$_{6}$
and Ba$_{2}$TmBiO$_{6}$ have been studied~\cite{Otsuka2015}. Besides
the basic thermodynamic and structural measurements at high temperatures,
very little information is known for these two materials. Thus, we
cannot extract much more physical understanding for the time being.
But these two materials remain as good candidates for frustrated FCC
systems with spin-orbit-entangled local moments~\cite{PhysRevB.95.085132}.

\subsection{General expectation for intrinsic quantum Ising magnets}

From our study of TmMgGaO$_{4}$ and the discussion on many other
Tm-based magnets, we think the intrinsic quantum Ising magnets can
widely exist in nature. The Tm$^{3+}$ ion in the D$_{3d}$ crystal
field is a bit special due to the high symmetry group of the D$_{3d}$
point group and the symmetry demanded crystal field singlets. In more
general cases~\cite{Chen2019intrinsic}, we do not have such a high
symmetry point group, and thus we think the intrinsic transverse field
can be more common in rare-earth magnets with lower crystal field
symmetries. To further remove the degeneracy, we have to get rid of
the Kramers' theorem. It is then interesting to search for the intrinsic
quantum Ising magnets among the rare-earth magnets with low crystal
field symmetries and integer-spin local moments.

\section*{Acknowledgments}

One of us (CJH) thanks Professor Youjin Deng for useful discussions
on QMC algorithm and providing computational resources supported by
the National Science Fund for Distinguished Young Scholars under Grant
No. 11625522. We also thank Kedar Damle for a conversation. 
This work is further supported by research funds from the Ministry of Science
and Technology of China with grant No.2016YFA0301001, No.2018YFGH000095
and No.2016YFA0300500, and from the Research Grants Council of Hong
Kong with General Research Fund Grant No.17303819. 
 
\appendix

\section{Results from linear spin-wave theory}

\label{append}

In this appendix, we provide the linear spin-wave theory and results
for the magnetic excitations in the three-sublattice magnetic orders.
The reason that we do this calculation is to clarify the discrepancy
between the number of the magnetic sublattices and the numbers of
the measured magnon branches.

For the three-sublattice magnetic ordered states, the system has $\sqrt{3}\times\sqrt{3}$
magnetic unit cell, each spin can be labeled by combination of magnetic
unit cell position ${\mathbf{r}}$ and sublattice index $s$ ($s=1,2,3$).
The mean-field ground-states can be obtained by Weiss mean-field theory,
where the mean-field spin orientations for each sublattices $s$ can
be labeled by unit vector $\mathbf{n}_{s}$. Then one can always associate
two unit vectors ${\mathbf{u}_{s}\cdot\mathbf{n}_{s}=0}$ and ${\mathbf{v}_{s}=\mathbf{n}_{s}\times\mathbf{u}_{s}}$
so that $\mathbf{n}_{s}$, $\mathbf{u}_{s}$ and $\mathbf{v}_{s}$
are orthogonal with each other. Next we perform Holstein-Primakoff
transformation for the spin operator $\mathbf{S}_{{\mathbf{r}}s}$,
\begin{eqnarray}
\mathbf{n}_{s}\cdot\mathbf{S}_{{\mathbf{r}}s} & = & S-b_{{\mathbf{r}}s}^{\dagger}b_{{\mathbf{r}}s}^ {},\\
(\mathbf{u}_{s}+i\mathbf{v}_{s})\cdot\mathbf{S}_{{\mathbf{r}}s} & = & ({2S-b_{{\mathbf{r}}s}^{\dagger}b_{{\mathbf{r}}s}})^{\frac{1}{2}}b_{{\mathbf{r}}s},\\
(\mathbf{u}_{s}-i\mathbf{v}_{s})\cdot\mathbf{S}_{{\mathbf{r}}s} & = & b_{{\mathbf{r}}s}^{\dagger}({2S-b_{{\mathbf{r}}s}^{\dagger}b_{{\mathbf{r}}s}})^{\frac{1}{2}}.
\end{eqnarray}
After performing Fourier transformation 
\begin{equation}
b_{{\mathbf{r}}s}=\sqrt{\frac{3}{N}}\sum_{\mathbf{k}\in{\overline{\text{BZ}}}}b_{\mathbf{k}s}e^{i\mathbf{R}_{{\mathbf{r}}s}\cdot\mathbf{k}},
\end{equation}
the spin Hamiltonian can be rewritten in terms of boson bilinears
as 
\begin{equation}
H_{\text{sw}}=\frac{1}{2}\sum_{\mathbf{k}\in{\overline{\text{BZ}}}}\Psi(\mathbf{k}){}^{\dagger}h(\mathbf{k})\Psi(\mathbf{k})+cosnt.,
\end{equation}
where
\begin{equation}
\Psi(\mathbf{k})=[b_{\mathbf{k}1},b_{\mathbf{k}2},b_{\mathbf{k}3},b_{\mathbf{-k}1}^{\dagger},b_{\mathbf{-k}2}^{\dagger},b_{\mathbf{-k}3}^{\dagger}]^{T}
\end{equation}
and $h(\mathbf{k})$ is a ${6\times6}$ Hermitian matrix, and $\overline{\text{BZ}}$
is the magnetic Brillouin zone. Then we can Bogoliubov diagonalize
$H_{\text{sw}}$ with ${\Psi(\mathbf{k})=T_{\mathbf{k}}\Phi(\mathbf{k})}$,
where 
\begin{equation}
\Phi(\mathbf{k})=[\beta_{\mathbf{k}1},\beta_{\mathbf{k}2},\beta_{\mathbf{k}3},\beta_{\mathbf{-k}1}^{\dagger},\beta_{\mathbf{-k}2}^{\dagger},\beta_{-\mathbf{k}3}^{\dagger}]^{T},
\end{equation}
is the diagonalized basis and $T_{\mathbf{k}}$ is the transformation
matrix. Details of diagonalization can be referred to \cite{petit2011,wallace1962,toth2015}.
The diagonalized Hamiltonian reads
\begin{eqnarray}
H_{\text{sw}} & = & \frac{1}{2}\sum_{\mathbf{k}\in\overline{\text{BZ}}}\Phi(\mathbf{k}){}^{\dagger}E(\mathbf{k})\Phi(\mathbf{k})+const.\nonumber \\
 & = & \sum_{\mathbf{k}\in\overline{\text{BZ}}}\omega_{\mathbf{k}s}\beta_{\mathbf{k}s}^{\dagger}\beta_{\mathbf{k}s}+const.,
\end{eqnarray}
where $E(\mathbf{k})=\mbox{diag}[\omega_{\mathbf{k}1},\omega_{\mathbf{k}2},\omega_{\mathbf{k}3},\omega_{-\mathbf{k}1},\omega_{-\mathbf{k}2},\omega_{-\mathbf{k}3}]$.
Within this formalism, we find that the coherent contribution to the
${\mathcal{S}}^{zz}$ correlator takes the following form

\begin{eqnarray}
 &  & {\mathcal{S}}^{zz}(\mathbf{k},\omega)\nonumber \\
 &  & \quad=\frac{S}{6}\sum_{s=1}^{3}[T_{\mathbf{k}}^{\dagger}\mathbf{U}^{z}(\mathbf{U^{z})}^{\dagger}T_{\mathbf{k}}]_{s+3,s+3}\delta(\omega-\omega_{-\mathbf{k}s}),
\end{eqnarray}
where $\mathbf{U}^{z}$ is a $6$-dimensional vector
\begin{eqnarray}
\mathbf{U}^{z} & = & [u_{1}^{z}+iv_{1}^{z},u_{2}^{z}+iv_{2}^{z},u_{3}^{z}+iv_{3}^{z},\nonumber \\
 &  & \,\,u_{1}^{z}-iv_{1}^{z},u_{2}^{z}-iv_{2}^{z},u_{3}^{z}-iv_{3}^{z}]^{T}.
\end{eqnarray}

Due to the quantum fluctuation, the magnetic orders are suppressed
from the mean-field values, as a result, the bandwidth of the 
single-magnon spectra will be renormalized. This is a well-known 
feature of the linear spin-wave theory~\cite{PhysRevB.79.144416}. 
If one is interested in more quantitative
features, one could use more involved renormalized spin-wave theory
that takes into account the suppression of the magnetic orders by
quantum fluctuations~\cite{PhysRevB.79.144416}. 
However, the linear spin-wave theory does provide
a useful understanding of the structure of the magnetic excitations.
In our spin-wave calculation, there are three branches of dispersions
that are consistent with the number of the magnetic sublattices.

 \bibliographystyle{apsrev4-1}
\bibliography{TMGO_PRX}
\end{document}